\documentclass[twocolumn]{aa}
\usepackage{graphics}
\usepackage{amsfonts}
\include{epsf}
\include{epstopdf}
\newcommand{\bfo}[1]{\mbox{\boldmath $#1$}}
\def\bvarphi{\mbox{\boldmath $\varphi$}}

\begin{document}
\newcommand{\beq}{\begin{equation}}
\newcommand{\eeq}{\end{equation}}
\def\la{\hbox{\raise.35ex\rlap{$<$}\lower.6ex\hbox{$\sim$}\ }}
\def\ga{\hbox{\raise.35ex\rlap{$>$}\lower.6ex\hbox{$\sim$}\ }}
\def\runit{\hat {\bf  r}}
\def\phunit{\hat {\bfo \bvarphi}}
\def\etaunit{\hat {\bfo \eta}}
\def\zunit{\hat {\bf z}}
\def\zetaunit{\hat {\bfo \zeta}}
\def\xiunit{\hat {\bfo \xi}}
\def\beq{\begin{equation}}
\def\eeq{\end{equation}}
\def\beqa{\begin{eqnarray}}
\def\eeqa{\end{eqnarray}}
\def\sub#1{_{_{#1}}}
\def\order#1{{\cal O}\left({#1}\right)}
\newcommand{\sfrac}[2]{\small \mbox{$\frac{#1}{#2}$}}
%
%
\title{{Low magnetic-Prandtl number flow configurations for cold astrophysical disk models: speculation and analysis}}

\author{O. M. Umurhan\inst{1,2}}

   \offprints{O.M. Umurhan \email{umurhan@maths.qmul.ac.uk}}

   \institute{
   Astronomy Unit, School of Mathematical Sciences, Queen Mary
   University of London, London E1 4NS, U.K.\
     \and
        Astronomy Department, City College of San Francisco,
      San Francisco, CA 94112, USA\
}

\date{}

 \abstract
     {Simulations of astrophysical disks in the shearing box that are subject to
 the magnetorotational instability (MRI) show that activity appears to be reduced
 as the magnetic Prandtl number P$_{{\rm m}}$ is lowered.  It is therefore important
 to understand the reasons for this trend, especially if this trend is shown
 to continue when higher resolution calculations are performed in the near future.
 Calculations for laboratory
 experiments show that saturation is achieved through
 modification of the background shear for P$_{{\rm m}} \ll 1$.}
      {Guided by the results of calculations appropriate for laboratory
  experiments when P$_{{\rm m}}$ is very low, the stability of   {inviscid} disturbances
  in a shearing box model immersed in a constant vertical background magnetic field
  is considered under a variety of shear profiles and boundary conditions in order
  to evaluate the hypothesis that modifications of the shear bring about saturation
  of the instability.
  Shear profiles $q$ are given by
  the local background Keplerian mean, $q_0$, plus time-independent
  departures, $Q(x)$, with zero average on a given scale.}
  {The axisymmetric linear stability of {inviscid} magnetohydrodynamic
  normal modes in the shearing box is analyzed.}
     {(i) The stability/instability of modes subject to modified shear profiles
    {may be interpreted} by a generalized Velikhov criterion given by an
  effective shear and radial wavenumber that are defined
  by the radial structure of the mode and the form of $Q$.
  (ii) Where channel modes occur,
  comparisons against marginally unstable disturbance in the classical
  case, $Q=0$,
  shows that
  all modifications of the shear examined here enhance mode instability.
  (iii) For models with boundary conditions mimicing laboratory experiments,
  modified shear profiles exist
  that stabilize a marginally unstable MRI for $Q=0$.
   (iv) Localized normal modes on domains of infinite radial extent characterized
   by either single defects or symmetric top-hat profiles for $Q$
   are also investigated.  If
   the regions of modified shear are less (greater) than the local Keplerian background, then
   there are (are no) normal modes leading to the MRI.}
     {The emergence and stability of the MRI is sensitive to the boundary conditions
  adopted.  Channel modes do not appear to be stabilized through modifications of the
  background shear whose average remains Keplerian.  However, systems that have non-penetrative
  boundaries can saturate the MRI through modification of the background shear.
  Conceptually equating the qualitative results from laboratory experiments to
  the conditions in a disk may therefore be misleading.}


\titlerunning{Speculation on low magnetic Prandtl number disk velocity profiles}


  \maketitle

\section{Introduction}
After almost 40 years of investigation the source
of the anomalous transport in accretion disks still
remains an open question.  As it is mostly assumed that
the transport is a reflection of some underlying turbulent state,
identification of its source mechanism and process has been
the focus of much research activity.  A leading
candidate mechanism is the magnetorotational
instability (Balbus \& Hawley 1991).  \par
The magnetic Prandtl number (P$_{\rm m}$), i.e. the
ratio of a fluid's viscosity to its magnetic diffusivity, appears
to play an important role in the ability of the MRI
to drive a fully-developed turbulent state.
Recently resolved simulations of the MRI in a shearing box
environment (e.g. Lesur \& Longaretti 2007; Fromang et al. 2008) show the appearance of a
turbulent state for moderately high Reynolds numbers (Re) when
P$_{\rm m}$ is an order $1$ quantity or higher.  However, the amount of
transport delivered by the MRI appears to depend on P$_{\rm m}$:
as this quantity decreases the vigor of the turbulent
state decreases and if P$_{\rm m}$ drops below
some critical value, then the turbulence appears to vanish altogether.
\par
Estimates of the properties of cold
astrophysical disks, like protoplanetary disks, show
that their characteristic P$_{\rm m}$'s are on the order of $10^{-6}$
at best and that the MRI may be operating only in the disk's
corona as its ionization fraction  is sufficient
to merit an MHD description there rather than near the disk midplane
(see Balbus \& Henri 2008 and references therein).
Thus, the fate of the MRI in disk environments in which
the magnetic Prandtl number is
characteristically small remains to be settled.
\par
It is not clear why turbulent transport appears to vanish
in numerical experiments when P$_{\rm m}$ weakens.
One possibility may be that under those conditions
the MRI cannot grow sufficiently to excite a secondary
transition that would, in turn, open the way to a turbulent cascade.
Theoretical considerations of the MRI for restrictive
configurations like laboratory experiments with cylindrical geometry predict that
a low  P$_{\rm m}$ sheared fluid undergoing the MRI
settles onto a pattern state whose momentum transport scales
as $\sim$ (P$_{\rm m}$)$^\alpha$ where $\alpha \ge 1$ -
in other words,  saturation of the instability
is sensitive to the microscopic viscosity of the fluid.
This counterintuitive result is rationalized by
noting that a weak fluid viscosity
means that it takes very little effort to readjust
the underlying flow profile and establish a new
shear in which
the instability cannot operate.  The stronger the viscosity, then,
it becomes more difficult for the fluid to shut off the shear
and the instability may operate unabatedly during the fluid's
nonlinear development and eventual cascade to turbulence
via secondary instabilities (e.g. parasitic instabilities, Xu \& Goodman, 1994).
\par
In the idealized quasi-linear
study of Knobloch \& Julien (2006) it was shown that
in the limit of very large hydrodynamic and magnetic Reynolds
numbers (i.e. Re $\gg 1$ and Re$_{\rm m} \gg 1$ respectively)
the azimuthal velocity profile tips over to shut off the instability
by establishing a new velocity profile which reduces the shear throughout the domain.
  {Their analysis shows that }
  {in the limits where Re $\rightarrow\infty$ and Re$_{\rm m} \rightarrow \infty $}
  {the resulting net velocity profile tends toward zero shear.}
Examinations
of the thin-gap Taylor-Couette system in the low P$_{\rm m}$ limit (Umurhan et al. 2007a; Umurhan et al. 2007b) show that once
the system reaches saturation,
the resulting velocity field within large mid-portions of the experimental domain is
characterized by weakened shear.  However these regions,
where the tendency toward instability is reduced, are sandwiched by regions of
strengthened shear, where the tendency toward instability is enhanced.  Nonetheless,
the aggregate configuration
results in a profile that is stable against the MRI or any secondary instability
and the system settles onto non-turbulent pattern state.  Equally relevant is that
in this pattern state
the amplitudes of all fluid quantities, except the modified shear,
scale as some power of the P$_{\rm m}$.  By comparison the
modification of the shear is an order 1 quantity.  Thus, in the
limit where P$_{\rm m}$ becomes very small only the modification to the background
shear appears as a noticeable response in the idealized studies of the laboratory setup.\par
\begin{figure}
\begin{center}
\leavevmode \epsfysize=5.cm
\epsfbox{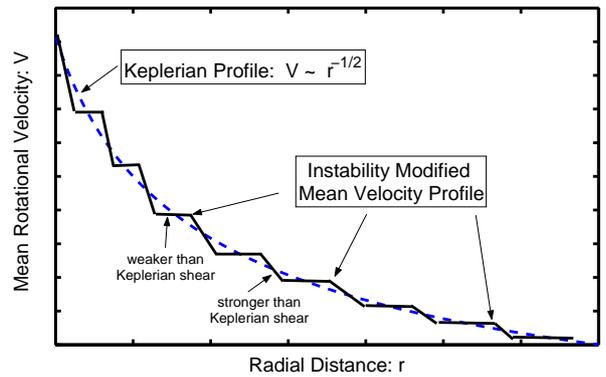}
\end{center}
\caption{Qualitative depiction of a hypothesized mean rotational velocity profile for
a low magnetic Prandtl number disk suggested by Jamroz et al. (2008).
The dashed line shows the usual Keplerian
velocity profile describing a rotationally supported
disk.  The mean velocity profile resulting
from the saturation of the MRI in the low P$_{\rm m}$ limit
is speculated to consist of alternating regions of
low and high shear (black lines). The mean shear over the
extent of the disk follows the general Keplerian profile.
}
\label{low_PM_figure1}
\end{figure}

Because resolved shearing box numerical experiments of high Re and low P$_{\rm m}$
flows \emph{appropriate for cold astrophysical disks} is currently out of reach,
speculation at this stage is justifiable.  In particular, what may be
the effective azimuthal velocity profile of high Re and low P$_{\rm m}$ disks (or their
sections)?
As Ebrahimi et al. (2009) recently note, quasi-linear saturation
of the shear profile might
be an unreasonable expectation given that the strong gravity
tends to restore the shear profile within disks
(cf. Julien \& Knobloch, review in preparation 2009).
However, the theoretical results of the
thin-gap Taylor-Couette system and the quasi-linear studies,
together with the objection of Ebrahimi et al.,  point to
a possible hybrid scenario:  What if the
velocity profile
of a low P$_{\rm m}$ disk is on average Keplerian but
locally exhibits alternating zones
of very weak (or no) shear  and very strong shear?  This hypothetical
configuration of the azimuthal velocity profile,
an example of which is depicted in Figure \ref{low_PM_figure1},  might very well be
driven into place by the MRI under very low P$_{\rm m}$ situations.
  {The recent studies of Julien \& Knobloch (2006) and  Jamroz et al. (2008) argue for the relevance of this scenario}.
\par
This study is an examination of the axisymmetric linear stability of incompressible
ideal magnetohydrodynamic disturbances
in a shearing box threaded by a constant vertical magnetic field
and characterized by a variety of mean velocity profiles, including an extreme
instance of the one appearing in Figure  \ref{low_PM_figure1}.
\footnote{{\em Extreme} in the sense that regions of sharp shear are represented
by delta-functions in the analysis herein.  See also discussion
in Section 6.}
 Such configurations
might be representative of real astrophysical disks characterized by
very small values of P$_{\rm m}$.\par
  {Guided by the low P$_{\rm m}$ results of Knobloch \& Julien (2005) and
Umurhan et al. (2007a-b),
the analysis
undertaken in this study is inviscid.  The only low P$_{\rm m}$ effect that is
 included here is the possibility that the
  fundamental shear profile is significantly modified.  This altered
 shear profile may be easily included into the equations of motion
  as any radially dependent barotropic steady shear profile is a permitted equilibrium flow
solution of the small shearing-box equations.}
  \par
The range
of questions that figure prominently are: (i) What happens to the unstable
mode leading to the MRI when there are departures of the steady
shear from the background Keplerian state? (iii) How does the nature
or existence of unstable modes depend upon the radial boundary conditions
of the shearing box (e.g. whether it is open or periodic in the
radial direction)? (ii) Can one infer a possible reason or explanation
for why turbulence seems to vanish in numerical experiments
of the shearing box when P$_{\rm m}$ is small?
\par
This work is organized as follows:  In Section 2 the equations of motion
are laid out and the framework within which the stability analysis is
discussed.  Section 3 develops the profiles for the steady state for given
modifications of the background shear while in Section 4 the normal mode stability
analysis is developed and the governing ODE is derived including a statement
of the generalized Velikhov criterion (Velikhov 1959).  Section 5 presents the
results of the stability analysis for a variety of boundary conditions
and shear profiles and their forms.  The implications of
the results are discussed in the final section.

\section{Equations of Motion}
The small shearing box equations (SSB, Lesur \& Longaretti 2007; Regev \& Umurhan 2008) are adopted
which are the incompressible incarnation of the usual shearing box
equations (Goldreich \& Lynden-Bell 1965).  Their usage
is justified as
compressibility is an inessential feature for the MRI.
The box is moving in a rotating frame moving with the local
Keplerian velocity $\bar V_{_K}$.\footnote{Overbars over quantities refer to their dimensional values.}
The length scales of the box are measured by a size $\bar L$ which is much smaller than
the radial scale of the disk $\bar R_0$.  On these scales, the background Keplerian shear
appears as a linear Couette flow profile.  Time units are scaled by
the local rotation time of the disk $\bar \Omega(\bar R_0)$, as measured at the
fiducial radius $\bar R_0$, implying that the velocities of interest
are scaled by $\bar U_0 = \bar L\bar \Omega(\bar R_0)$.
Given this, the disparity of the ratio of the
local disk soundspeed to the Keplerian speed (since the gas is relatively cold),
and the assumption that the scale $\bar L$ is much less than the vertical scale
height of the disk,
the flow dynamics
have the character of incompressible rotating magnetic Couette flow.
Magnetic fields are
scaled by a reference field scale $\bar B_0$ and the density is $\bar \rho$.  These
quantities form a velocity defined
to be the Alfven speed via $\bar U_{_A}^2 \equiv \bar B_0^2/4\pi\bar \rho$.  Furthermore
the Cowling number is defined
\[
{\cal C} \equiv \frac{\bar U_{_A}^2}{\bar U_0^2},
\]
  {which may be considered as related to the inverse-square of the ``$\beta$-parameter" of plasma physics.  This is not to be confused with the parameter $\beta$ used later on in this work to designate radial wavenumbers.}
The non-dimensionalized
SSB equations of motion
are
\beqa
(\partial_t + {\bf u}\cdot\nabla) {\bf u} + 2\Omega_0 {\hat {\bf z}}\times {\bf u}
&=& \nonumber \\
-\frac{1}{\rho}\nabla p &+& 2\Omega_0^2 q_0 x {\hat {\bf x}} +
{\cal C}{\bf J}\times {\bf B},
 \\
\nabla\cdot {\bf u} &=& 0, \\
\partial_t {\bf B} &=& \nabla \times {\bf u} \times {\bf B}, \label{B_eqn}\\
{\bf J} &=& \nabla \times {\bf B},
\eeqa
in which ${\bf u}$ is the velocity vector, $p$ is the pressure, $\rho$ is the
constant density (which is 1 in these units).
$\Omega_0$ is the local rotation rate but in the units
used here it is also $1$.  The parameter $q_0$ defines the
amplitude of the local Keplerian shear and is formally given
by
\[
q_0 \equiv -\left(\frac{\bar R}{\bar\Omega}\cdot\frac{\partial \bar \Omega}{\partial \bar R}\right)_{\bar R_0},
\]
which is equal to 3/2 for rotationally supported Keplerian disks.
  {This is the value assumed for $q_0$ throughout the remainder of this work.}
The magnetic field is given by the vector ${\bf B}$ and the
current ${\bf J}$ and is governed by Ampere's Law.
\footnote{Note that by its construction (\ref{B_eqn})
contains the statement that the time evolution of the divergence
of the magnetic field is zero. If initially
$\nabla\cdot{\bf B} = 0$ everywhere, then it remains identically
zero subsequently.}
\section{Steady State}
It is assumed that the fluid is of constant density $\rho_0$
and the fluid is threaded uniformly throughout by a constant
vertical magnetic field, i.e. ${\bf B} = B_z {\hat {\bf z}}$.  The current
source of this background
field is taken to be external to the domain.  As stated in the Introduction
the steady velocity field that represent departures to the Keplerian
shear are taken as given and represented by $U(x)$.  The total
mean velocity is written as $V(x)$ which is the sum of the Keplerian
portion $-q_0\Omega x$ and the departures $U$.
Thus
the steady state configuration requires satisfying the radial
momentum balance which becomes
\beq
-2\Omega_0 (-q_0\Omega_0 x + U) = -\partial_x \varpi_0 + 2\Omega_0^2 q_0 x,
\eeq
in which the total mechanical and magnetic pressure is given by
\[
\varpi_0 = p_0/\rho_0 + \sfrac{1}{2}{\cal C} B_z^2,
\]
where $p_0$ is the steady state mechanical pressure field.  For the
model analyzed here where $B_z$ is constant,
the steady pressure field $p_0$ is related to
the azimuthal velocity departures by
\beq
\partial_x p_0 = 2\Omega_0 U \bar\rho. \label{radial_geostrophy}
\eeq
The above expression represents a radially geostrophic state.  Note
that the solutions to $p_0$ are smooth and well behaved so long
as $\tilde V$ is continuous across the domain.
  {In the axisymmetric linear stability analysis that follows
only the steady state total vorticity field, denoted by $q$, plays a role
and it is defined as}
\beq
q = q_0 + Q,
\eeq
where $Q(x) \equiv -\partial_x U/\Omega_0$.
A variety of functional forms for $Q$ will be examined.  The only
restriction on $Q$ will be that its domain integral is bounded, i.e.,
\[
\left|\int_{{\cal D}} Q dx\right| < \infty,
\]
where ${\cal D}$ symbolizes the domain under consideration.
In many cases this integral will be zero.

\section{Linear Theory}
  {Axisymmetric infinitesimal disturbances about the steady state shaped by $q(x)$
are introduced into the governing equations
of motion (1-4) revealing,}
\beqa
\partial_t u' - 2\Omega_0 v' &=& -\partial_x \left(p'/\rho_0 + {\cal C} B_z b_z'\right)
+ {\cal C} B_z \partial_z b_x', \label{mhd_u_eqn}\\
\partial_t v' + (2-q)\Omega_0 u' &=&  {\cal C} B_z \partial_z b_y', \label{mhd_v_eqn}\\
\partial_t w' &=& -\partial_z \left(p'/\rho_0 + {\cal C} B_z b_z'\right)
+ {\cal C} B_z \partial_z b_z, \label{mhd_w_eqn} \\
\partial_x u' + \partial_z w' &=& 0, \label{mhd_inc}\\
\partial_t b_x' &=& B_z \partial_z u, \label{b_x_eqn}\\
\partial_t b_y' &=& - \Omega_0 q b_x + B_z \partial_z v', \label{b_y_eqn}\\
\partial_t b_z' &=& B_z \partial_z w', \label{b_z_eqn}
\eeqa
in which $u',v' w'$ and $p'$ denote perturbations in the
radial, azimuthal vertical velocities and pressures, respectively, while $b'_{x,y,z}$
represent perturbations in the magnetic field components.
  {It should be recalled that $q(x)$ is unspecified at this stage.  However the
existence of a steady state is assumed for a given reasonable $q(x)$ profile.  }
Normal mode perturbations are assumed which are periodic in the vertical direction.
A perturbation variable $f'(x,z,t)$ is thus expressed via the ansatz  $f'(x) \exp(\sigma t + ikz)$
where $\sigma$ is the temporal response while $k$ is the vertical
wavenumber.   The incompressibility condition  (\ref{mhd_inc})
means that a streamfunction $\psi'$ may be defined such that,
\[
u' = \partial_z \psi', \qquad w' \equiv -\partial_x \psi'.
\]
Similarly the expression (\ref{b_x_eqn}) and (\ref{b_z_eqn})
allow for the definition of a flux function $\Phi$ in which
the radial and vertical components of the perturbed magnetic
field are expressed via
$
b_x = \partial_z \Phi'$ and $ b_z \equiv -\partial_x \Phi'$.

Utilizing these definitions,
the linearized perturbation equations
are reduced to a single one for the streamfunction (wherein and henceforth
the primes are dropped),
\beqa
\partial_x^2\psi - \kappa^2\psi &=& 0,
\label{MHD_psi_eqn}
\eeqa
in which
\beqa
\kappa^2 &\equiv& k^2\left(1+\frac{
\omega^2\sigma^2
- \Omega_0^2 2q \omega_{az}^2
}{
(\sigma^2 + \omega_{az}^2)^2}
\right); \quad \omega_{az}^2 \equiv {\cal C}B_z^2 k^2,
\label{kappa_definition}
\eeqa
where $\omega_{az}$ is the Alfven frequency
and the spatially dependent
epicyclic frequency is defined by $\omega(x)^2 \equiv
2(2-q)\Omega_0^2$.
For later usage $\kappa$ is rewritten
in the alternate form,
\beq
\kappa^2 = k^2\left(1+\frac{
4\Omega_0^2\sigma^2}
{(\sigma^2 + \omega_{az}^2)^2}
-\frac{2\Omega_0^2 q}{\sigma^2 + \omega_{az}^2}
\right).
\label{kappa_definition_alternate}
\eeq
The solutions
to this system is governed by several parameters including the vertical
wavenumber $k$, the Cowling number, ${\cal C}$,   {the radial size of the domain
(appearing below) $L$ and the form and amplitude of the deviation
shear profile $Q$.}  Taken as a representation of a small disk section
it is assumed that there exists a minimum vertical
wavenumber $k_0$ for disturbances since it is not reasonable to
consider fundamental disturbances whose vertical extent is much
larger than the disk height itself (Curry et al. 1994; and see Discussion).
\par
Before proceeding to these results it is important to keep in
mind a number of matters.    {First,
Knobloch (1992) showed that for vertical field configurations
like that considered here, together with boundary conditions in which
individual disturbances or their combinations are zero on the boundaries,
the character of the \emph{normal mode
response} of the MRI is that  $\sigma$ is either real (growing/decaying
exponential modes) or imaginary (oscillating modes), but never complex}.\footnote{ It was
also shown in that study that complex
values of $\sigma$ can only occur if the field configuration is
helical. In this case the instability is of a traveling wave.}
A version of the argument leading to this
conclusion is given in Appendix \ref{Integral_Argument}.
Thus when  the character of the fluid response is analyzed,
 $\sigma^2$ will be taken as $\in$ Reals.   The relevance
 of this observation is that
 as a function
 of the system's parameters, a mode becomes unstable by passing
 through $\sigma^2 = 0$,
 sometimes known as
  {\em exchange of stabilities} (Chandrasekhar 1961).
 As such, in some instances
 the stability analysis
 of this system will focus on identifying which mode
 becomes marginal (i.e. $\sigma^2 = 0$) and under which parameter conditions.  \par
 Secondly, the very same integral argument developed in
 Appendix \ref{Integral_Argument} also leads to a general
 statement about the relationship between the temporal response
 and the structure of the modes in the domain,
 \beq
 1 + \frac{\bar\beta^2}{k^2}
 + \frac{4\Omega_0^2 \sigma^2}{(\sigma^2 + \omega_{az}^2)^2}
 -\frac{2\Omega_0^2 \bar q}{\sigma^2 + \omega_{az}^2} = 0.
 \label{genA}
 \eeq
  $\bar\beta$ is like an average radial wavenumber of the disturbance,
 \beq
 \bar\beta^2 \equiv
 \frac{\int_{{\cal D}}|\partial_x\psi|^2 dx}
 {\int_{{\cal D}}|\psi|^2 dx},
 \label{genB}
 \eeq
 and, evidentally, $\bar\beta^2$ is always greater than zero.
 $\bar q$ represents the mode weighted average
 shear over the domain
 \beq
 \bar q \equiv q_0 + \frac{\int_{{\cal D}}Q|\psi|^2 dx}
 {\int_{{\cal D}}|\psi|^2 dx}.
 \label{genC}
 \eeq
 One immediate consequence of this is that if $\bar q = 0$
(i.e. the
 mode weighted average shear over the domain is zero),
 then there is no possibility of instability.  This follows from
  evaluating the criterion for marginality in (\ref{genA}),
 setting $\sigma^2$ to zero and revealing
 \beq
 1 + \frac{\bar\beta^2}{k^2}
 -\frac{2\Omega_0^2 \bar q}{\omega_{az}^2} = 0.
 \eeq
 The above expression cannot be satisfied if $\bar q$ is zero.  In general
 this relationship is not useful for direct calculations
   {as the quantities therein depend implicitly upon the
 eigenvalue $\sigma$.}
 However, it will be called upon to aid in interpreting
 the results that are obtained from specific direct
 calculations in Section 6.

\section{Results}\label{Results_Section}
\subsection{Classical MRI modes and Channel Solutions: A Review }\label{classical_limit_theory}
The classical system is recovered for $Q=0$.  This means that
$\kappa$ is independent of $x$ and
\beq
\kappa^2 \rightarrow \kappa_0^2 \equiv
 k^2\left(1+\frac{
4\Omega_0^2\sigma^2}
{(\sigma^2 + \omega_{az}^2)^2}
-\frac{2\Omega_0^2 q_{0}}{\sigma^2 + \omega_{az}^2}\right).
\label{kappa_0_def}
\eeq
Solutions of (\ref{MHD_psi_eqn}) are sought on a periodic
domain $L$ and are given in general by
\beq
\psi = A\cosh(\kappa_0 x + \phi),
\eeq
where $A$ and $\phi$ are constants set by the boundary conditions of the system.
The periodicity condition that $\psi(x+L/2) = \psi(x-L/2)$ then means that
\beq
1-\cosh\left[L\kappa_0\right]=0.
\eeq
which has the solution
$\kappa_0L = 2 n i \pi$ where $n$ is any integer including zero.  This condition recovers
the incompressible limit of the classical dispersion relationship
of the ideal case (e.g. Acheson \& Hide 1973; Balbus \& Hawley 1991) where
specifically,
\beq
\left(\beta_n^2 + k^2\right)(\sigma^2 + \omega_{az}^2)^2
+\omega_0^2k^2\sigma^2 - 2\Omega_0^2 q_0 k^2\omega_{az}^2 = 0.
\label{dispersion_condition_uniform_q}
\eeq
in which $\omega_0^2 \equiv 2(2-q_0)\Omega_0^2$ is the epicyclic frequency
and where the radial wavenumber $\beta_n$ is
\[
\beta_n^2 \equiv \frac{4n^2\pi^2}{L^2}.
\]
The solutions
for the temporal response, written in terms of respected pairs,
is given by
\beqa
& & \sigma^2 = \sigma_{_{0n}}^2(\beta_n^2) \equiv -\left(\omega_{az}^2 + \frac{k^2}{\beta_n^2 + k^2}\frac{\omega_0^2}{2}\right) \nonumber \\
& &  \ \pm\left[
\left(\omega_{az}^2 + \frac{k^2}{\beta_n^2 + k^2}\frac{\omega_0^2}{2}\right)^2
-\omega_{az}^2\left(\omega_{az}^2 - \frac{2\Omega_0^2 q_0 k^2}{\beta_n^2 + k^2}\right)
\right]^{1/2}. \ \
\label{general_temporal_response_classic_MRI}
\eeqa
The solutions associated with the `-' branch are termed the hydrodynamic inertial (HI) modes
while those associated with the `+' are the hydromagnetic inertial (HMI) modes
(viz. Acheson \& Hide 1973).  For a given mode $n$, the magnetorotational instability occurs
for the latter of these provided that $\omega_0^2 > 0$ and
\beq
\omega_{az}^2 - \frac{2\Omega_0^2 q_0 k^2}{\beta_n^2 + k^2} < 0.
\label{criterion_for_instability_classic_mri}
\eeq
The background field must be sufficiently weak for the HMI-modes
to be unstable.
\par
For given values of $\omega_{az}^2$ the absolute
minimum criterion required for instability to set in (i.e. for the mere possibility
of $\sigma^2 > 0$) is when $n=0$, i.e. $\kappa_0 = 0$, and
the condition in (\ref{criterion_for_instability_classic_mri}) predicts this to happen
if,
\beq
2\Omega_0^2 q_0 - \omega_{az}^2 > 0,
\label{velikhov_condition}
\eeq
(Velikhov 1959; Acheson \& Hide 1973).  This particular special
mode, in which its radial structure is uniform,
is generally referred to in the literature as the
\emph{channel mode} which describes purely 2D flow with no
vertical velocity.  The temporal response in this case is
given by $\sigma^2_{_{0n}} = \sigma_{_{00}}^2$ in which,
\beqa
& & \sigma_{_{00}}^2 = -\left(\omega_{az}^2 + \frac{\omega_0^2}{2}\right) \nonumber \\
& & \ \ \   \pm\left[
\left(\omega_{az}^2 + \frac{\omega_0^2}{2}\right)^2
-\omega_{az}^2\left(\omega_{az}^2 - {2\Omega_0^2 q_0 }\right)
\right]^{1/2}.
\label{sigma_0_sol}
\eeqa
When unstable (for the HMI-mode branch) this radially uniform
mode plays the central role in the development
of MRI induced turbulence in numerical experiments.
If $2-q_0 > 0$, then the
expression for $\sigma_{_{00}}^2$ is
\beqa
& & \sigma_{_{00}}^2 + \omega_{az}^2 =
-\Omega_0^2(2-q_0) \nonumber \\
& & \ \ \  \pm\Omega_0^2(2-q_0)\left[
1 +  \frac{4\omega_{{az}}^2}{\Omega_0^2(2-q_0)}
+ \frac{8\Omega_0^2 q_0\omega_{{az}}^2}{\Omega_0^4(2-q_0)^2}
\right]^{1/2}.
\label{classical_channel_mode}
\eeqa
The combination expression $\sigma_{_{00}}^2 +   \omega_{{az}}^2$
(which shall appear on a number of occasions in the following discussion)
is always positive for HMI modes and is always negative for the HM modes
since the expression inside the square-root operation is always positive.
\par
It should be noted that for $Q=0$ there are no \emph{localized} normal modes possible
for a domain which is infinite in extent.  Put in another way, there is no
solution of (\ref{MHD_psi_eqn}) for modes on an infinite domain with both (a)
all quantities going to zero as $x \rightarrow \pm \infty$ and (b)
 $\kappa$ is constant as given in (\ref{kappa_0_def}).
\subsection{Weak shear variations on a finite domain: $0< Q \ll 1$}\label{small_Q_theory}
  {In this section
weak shear variations are considered whose average are zero on the domain $L$.
Before proceeding a caveat ought to be stated.
  Despite the spatially periodic
nature of the shear $Q$, the analysis necessary to determine the temporal response
uses a multiple time-scale analysis.  When periodic boundary conditions
are imposed, as they are in Section 5.2.3, the multiple time-scale analysis
is akin to a Floquet analysis.  In this instance a long-time scale behavior
(in the form of correspondingly weak corrections to the temporal response $\sigma$) must be
invoked in order to ensure that the resulting perturbation solutions
remain periodic on the length scale $L$.  For boundary conditions other than
periodic ones (Sections 5.2.1 and 5.2.2),
the multiple-time scale analysis employed is a generic procedure
involving perturbation series expansions and imposition of solvability conditions
at successive perturbation orders (Bender \& Orszag 1999).  The role of the
solvability conditions is to make sure that higher order perturbation solutions
satisfy the boundary conditions of the system.}
\par
The shear with zero average on the domain of length $L$
may be decomposed into the Fourier series expansion
\beq
Q = \epsilon Q_1; \qquad Q_1 = \sum_{n=1}^{\infty}{
q_n \sin\left(\frac{2n\pi}{L}x-\frac{\pi}{2}\right)},
\label{small_Q_shear_profile}
\eeq
where the parameter $\epsilon \ll 1$ measures the overall severity of the
shear. $q_n$ measures the Fourier amplitude of the shear component $n$
and is treated hereafter as a tunable parameter.  Reference to
(\ref{radial_geostrophy}) indicates that this form
for $Q$ ensures that the deviation steady velocity $U$ and, hence, the
radial pressure gradients are zero
at the boundaries $x=\pm L/2$.
The governing equation, with $Q$ as given above, is expressed as
\beq
\partial_x^2\psi - \kappa_0^2\psi = -\epsilon\frac{2\Omega_0^2 k^2 Q_1}{\sigma^2 + \omega_{az}^2}\psi.
\eeq
Solutions of this are developed in a singular perturbation
series expansion in powers of $\epsilon$.
Thus the ansatz,
\beqa
\psi &=& \psi_0 + \epsilon\psi_1 + \epsilon^2 \psi_2 + \cdots \nonumber \\
\sigma^2  &=& \sigma^2_0 + \epsilon\sigma^2_1 + \epsilon^2\sigma^2_2 + \cdots \nonumber \\
\kappa_0^2 &=& \kappa^2_{_{00}} + \epsilon \kappa^2_{_{01}} + \epsilon^2 \kappa^2_{_{02}}
+\cdots
\eeqa
Given the definition of $\kappa_0^2$ given in (\ref{kappa_0_def}) it follows that
$\kappa_{_{00}}^2 = \kappa_0^2(\sigma_0^2)$ and
\beqa
 \kappa^2_{_{01}} &\equiv& \left(\frac{\partial\kappa_0^2}{\partial\sigma^2}\right)_{\sigma^2_0} \sigma^2_1,
 \nonumber \\
 \kappa^2_{_{02}} &\equiv&  \frac{1}{2}
 \left(\frac{\partial^2\kappa_0^2}{\partial\sigma^4}\right)_{\sigma^2_0} \sigma^4_1
 +\left(\frac{\partial\kappa_0^2}{\partial\sigma^2}\right)_{\sigma^2_0} \sigma^2_2,
\eeqa
wherein
\beq
\left(\frac{\partial\kappa_0^2}{\partial\sigma^2}\right) =
\frac{k^2}{\sigma^2 + \omega_{az}^2}
\left(-\frac{\omega_0^2}{\sigma^2 + \omega_{az}^2} + \frac{8\Omega_0^2\omega_{az}^2}{(\sigma^2 + \omega_{az}^2)^2}
\right).
\eeq
  {The possibility that  $\sigma_1^2 = 0$ must be allowed (see below) This is
why the next order term is also included in the expansion.}
With these expansions in mind (\ref{MHD_psi_eqn}) is written out for each
power of $\epsilon$.  To lowest order
\beq
\partial_x^2\psi_0 - \kappa_{_{00}}^2\psi_0 =0, \label{small_Q_order_0}
\eeq
and to order $\epsilon$
\beq
\partial_x^2\psi_1 - \kappa_{_{00}}^2\psi_1 = \left(\kappa_{_{01}}^2
-\frac{2\Omega_0^2 k^2 Q_1}{\sigma_0^2 + \omega_{az}^2}\right)\psi_0. \label{small_Q_order_1}
\eeq
The order $\epsilon^2$ form is relegated until later.  Solutions to
(\ref{small_Q_order_0}-\ref{small_Q_order_1}) are developed in the following
subsections for various boundary conditions and restrictions at $x=\pm L/2$.
\subsubsection{Channel Conditions}\label{ChannelWalls_smallQ}
Channel conditions require that the perturbed radial velocity is zero at
the boundaries $x=\pm L/2$.  This, in turn, amounts to requiring that $ik\psi = 0$ at these
positions, or in terms of the perturbation variables
\[
\psi_0 = \psi_1 = 0, \qquad {{\rm at}} \ \ x = \pm {L}/{2}.
\]
The lowest order solution is given by
\beq
\psi_0 = \sum_{m=1}^{\infty} \psi_{0m} =  \sum_{m=1}^{\infty}A_m \sin \gamma_m(x+L/2);
\quad \gamma_m \equiv \frac{m \pi}{L},
\label{Channel_Conditions_small_Q_theory_psi_solution}
\eeq
and where $\kappa_{00}^2 = -\gamma_m^2$.    {The general solution for $\psi_0$ is
written out as a sum of the linearly independent solutions $\psi_{0m}$.  The index
$m$ ought not be confused with the index $n$ which refers to the Fourier component
of the shear profile $Q$ appearing in (\ref{small_Q_shear_profile}).
Stability, on the other
hand, is evaluated by examining each individual eigenmode $m$.
Thus all eigenvalues $\sigma$ (and their perturbative corrections)
will be identified according to which eigenmode under examination,
e.g. $\sigma_{0m}, \sigma_{1m}$ etc.}
  {Note also that here the index $m$ does not include $m=0$
as this would correspond to the trivial state $\psi = 0$.}\par
Given the functional form of $\kappa_0^2$ found in
(\ref{kappa_0_def}), these relationships imply that $\sigma_{0}^2$
is equal to the functional form $\sigma_{0n}^2$ found in
(\ref{general_temporal_response_classic_MRI}) except with $\beta_n$ replaced by $\gamma_m$,
i.e. $\sigma^2_0 = \sigma_{0n}^2(\gamma_n^2)$.
  {The stability of these modes is now a function of the mode number $m$ and, henceforth,
the summation sign in (\ref{Channel_Conditions_small_Q_theory_psi_solution})
will be dropped and it shall be henceforth understood that
when reference is made to $\psi_0$, a particular eigenmode mode $m$ will be analyzed.}
At the next order it can be seen that in order to have
solutions that satisfy the boundary condition that $ik\psi_1 = 0$ at the two endpoints,
the terms on the RHS of (\ref{small_Q_order_1})
must satisfy a particular solvability condition which will relate the correction $\sigma_1$
to the variation $Q_1$.  This is seen by multiplying
(\ref{small_Q_order_1}) by $\psi_0$
given in (\ref{Channel_Conditions_small_Q_theory_psi_solution}) and integrating the
result across the domain from $-L/2$ to $L/2$.\footnote{This procedure of applying
a solvability condition actually requires multiplying across by the
adjoint solution of $\psi_0$.  However since the eigenfunctions comprising $\psi_0$
and the linear operator of (\ref{small_Q_order_0}) are real and Hermitian, the adjoint
is $\psi_0$.}
In a general sense, then, one finds,
\beq
\kappa_{_{01}}^2 =
\left(\frac{\partial\kappa_0^2}{\partial\sigma^2}\right)_{\sigma^2_{0m}} \sigma^2_{1m} =
\frac{2\Omega_0^2 k^2}{\sigma_{0m}^2 + \omega_{az}^2}\frac{<Q_1\psi_{0m}^2>}{<\psi_{0m}^2>},
\eeq
in which the bracket notation is
\beq
<\diamond> \equiv \int_{-L/2}^{L/2}{\diamond} \ \ dx.
\eeq
It follows that
\beq
\sigma_{1m}^2 = -q_m(-1)^m \Upsilon(m),
\label{sigma_1_Channel_Wall_sol}
\eeq
in which
\[ \Upsilon(m) \equiv
\frac{\Omega_0^2 (\sigma_{0m}^2 + \omega_{az}^2)^2}
{-\omega_0^2\sigma_{0m}^2 + 2\Omega_0^2(2+q_0)\omega_{az}^2}.
\]
 The denominator
term of $\Upsilon(m)$ never crosses zero
for physically relevant values of the parameters so that $\Upsilon > 0$.\par
\begin{figure}
\begin{center}
\leavevmode \epsfysize=6.2cm
\epsfbox{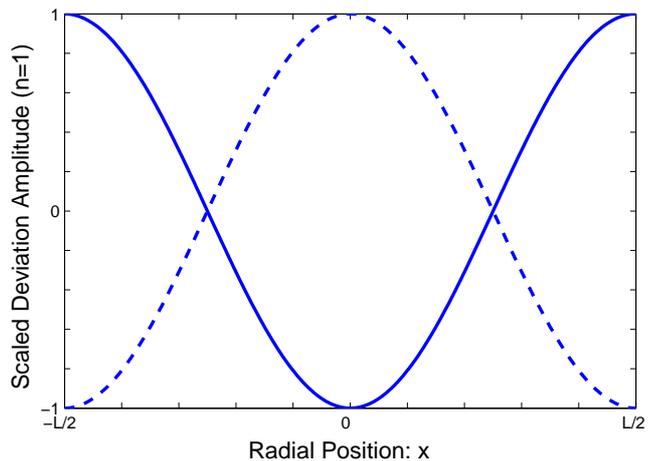}
\end{center}
\caption{The $n=1$ component of the shear profile (\ref{small_Q_shear_profile}).  The solid
line $q_1 = 1$ and the dashed line $q_1 = -1$.  For problems in which channel
wall conditions are imposed the solid line corresponds to a stable deviation profile while
the dashed line corresponds to an unstable profile.  The situation is reversed
when fixed-pressure boundary conditions are imposed. }
\label{small_Q_profile}
\end{figure}

For illustration consider the $m=1$ mode (which is also the most unstable one)
near marginality in the limit where
$kL$ is very large.  In order for $\sigma_{01}^2$ to be nearly zero
it must be that $\omega_{az}^2 \approx 2\Omega_0^2 q_0$, cf. (\ref{velikhov_condition}).
  {It follows that the correction for this mode, $\sigma_{11}$, is given approximately by
the solution to}
\beq
\sigma_{11}^2 \approx q_1 \frac{\omega_{az}^2}{2(2+q_0)} = q_1 \frac{\Omega_0^2q_0}
{(2+q_0)},
\eeq
  {where the last equality is established because the marginal mode, $\sigma_{01}^2 \approx 0$
is under examination.}
The introduction of this profile pattern stabilizes this   {otherwise marginal} mode
 if $q_1 < 0$
and destabilizes it when $q_1 > 0$.   The corresponding shear profile for this example is depicted in
Figure \ref{small_Q_profile}.  It can be seen that the modified shear is stabilizing when the
shear is positive near the boundaries and negative in the interior.

\subsubsection{Pressure Conditions}\label{small_Q_pressure_conditions}
An analysis of (\ref{mhd_w_eqn}) and (\ref{b_z_eqn}) shows that
the total magnetic plus mechanical pressure perturbations are proportional
to the radial gradient of $\psi$ i.e.,
\beq
p/\rho_0 +  B_z{\cal C} b_z = \frac{\sigma^2 + \omega_{az}^2}{\sigma}\partial_x \psi.
\eeq
Fixing the total pressure at the boundary in a Lagrangian sense (Curry et al. 1994) is
equivalent to setting to zero the above expression
at $x=\pm L/2$ which, in the present case, means setting $\partial_x \psi = 0$ at those
locations.
\footnote{Note that this is valid since by construction the radial
gradient of the steady state pressure field is zero at the end points.}
The solution $\psi_{0m}$ is then given by
\beq
\psi_{0m} = A_m \cos \gamma_m(x+L/2);
\quad \gamma_m \equiv \frac{m \pi}{L},
\label{Pressure_Conditions_small_Q_theory_psi_solution}
\eeq
which is very similar in form to (\ref{Channel_Conditions_small_Q_theory_psi_solution}) except
that the solution in the domain is composed of even functions and the series
includes the mode with $m=0$, which is the classical channel mode.  The correction to the
temporal response for all values of $m\neq 0$ is
\beq
\sigma_{1m}^2 = q_m(-1)^m \Upsilon(m),
\label{sigma_1_fixed_pressure_sol}
\eeq
which is the negative of the solution found in (\ref{sigma_1_Channel_Wall_sol}).
  As Figure \ref{small_Q_profile}
indicates,
  this result says that shear profiles
in a channel configuration
that promote stability (instability) play a destabilizing (stabilizing)
role for configurations with fixed-pressure boundary conditions .\par
However the $m=0$ mode is a special case which requires a separate analysis.  In its
simplest guise, the channel mode $\psi_{00}$ is a constant with respect to $x$.  Because
of the periodicity of $Q_1$ it means that $<Q_1\psi_{00}^2> = 0$ and the analysis
advanced up until this point to evaluate the correction $\sigma_1$ shows that it
will be zero at this order.  Thus given this and the fact that $\kappa_{_{00}}^2 = 0$
the solution must be expanded according to,
\beqa
\psi_0 &=& \psi_{00}  +\epsilon \psi_{10}+  \epsilon^2 \psi_{20} + \cdots \nonumber \\
\sigma^2  &=& \sigma^2_{00} + \epsilon^2\sigma^2_{20} + \cdots \nonumber \\
\kappa_0^2 &=&  \epsilon^2 \left(\frac{\partial\kappa_0^2}{\partial\sigma^2}\right)_{\sigma^2_{00}}\sigma^2_{_{20}}
\eeqa
\beq
\left(\frac{\partial\kappa_0^2}{\partial\sigma^2}\right) =
\frac{k^2}{\sigma^2_{_{00}} + \omega_{az}^2}
\left(-\frac{\omega_0^2}{\sigma^2_{00} + \omega_{az}^2} + \frac{8\Omega_0^2\omega_{az}^2}{(\sigma^2_{_{00}} + \omega_{az}^2)^2}
\right),
\eeq
where $\sigma^2_{_{00}}$ the response of the channel mode is given in (\ref{sigma_0_sol}).  The equations for the solution of $\psi$ are in successive order given
by
\beqa
\partial_x^2 \psi_{00} &=& 0, \\
\partial_x^2\psi_{10} &=&
-\frac{2\Omega_0^2 k^2 Q_1}{\sigma_{_{00}}^2 + \omega_{az}^2}\psi_{00}, \\
\partial_x^2\psi_{20} &=& \left(\frac{\partial\kappa_0^2}{\partial\sigma^2}\right)_{\sigma^2_{_{00}}} \sigma^2_{20} \psi_{00}
-\frac{2\Omega_0^2 k^2 Q_1}{\sigma_{_{00}}^2 + \omega_{az}^2}\psi_{10}.
\label{psi_20_eqn}
\eeqa
The first of these says that the lowest order solution which satisfies
the boundary condition is a constant $\psi_{00} = A_{00}$.  The next
order down the solution is
\beq
\psi_{10} =
-\frac{2\Omega_0^2 k^2}{\sigma_{_{00}}^2 + \omega_{az}^2}{\cal Q} A_{00},
\label{psi_10_sol}
\eeq
where
\beqa
& & {\cal Q} = - \sum_{n=1}^{\infty}
q_n \left(\frac{L}{2n\pi}\right)^2\Biggl[
\sin\left(\frac{2n\pi}{L}x + \theta_n\right) \nonumber \\
& & \hskip 4.0cm  -\frac{2n\pi}{L}x(-1)^n\cos \theta_n \Biggr].
\eeqa
In other words the solution ${\cal Q}$ is such that $\partial_x^2 {\cal Q} = Q_1$
which satisfies the boundary condition $\partial_x{\cal Q}|_{x = \pm L/2} = 0$.
In order to generate a bounded solution for $\psi_{20}$ a solvability condition must
be applied to the RHS of (\ref{psi_20_eqn}), which is obtained by re-expressing
$\psi_{10}$ in terms of its solution as given in (\ref{psi_10_sol})
and then integrating
the equation from across the domain and setting the result to zero:
\[
\left(\frac{\partial\kappa_0^2}{\partial\sigma^2}\right)_{\sigma^2_{_{00}}} \sigma^2_{20}
= -\left(\frac{2\Omega_0^2 k^2}{\sigma_{_{00}}^2 + \omega_{az}^2}\right)^2<Q_1 {\cal Q}> A_{00}.
\]
However, given the relationship between ${\cal Q}$ and $Q_1$   {it follows after an integration
 by parts that in fact},
\beq
\left(\frac{\partial\kappa_0^2}{\partial\sigma^2}\right)_{\sigma^2_{00}} \sigma^2_{20}
= \left(\frac{2\Omega_0^2 k^2}{\sigma_{00}^2 + \omega_{az}^2}\right)^2< (\partial_x{\cal Q})^2> .
\eeq
Given that for HMI modes
\[
\left(\frac{\partial\kappa_0^2}{\partial\sigma^2}\right)_{\sigma^2_{00}} > 0,
\]
the above solvability condition says something remarkable: that \emph{the
channel mode is always destabilized} no matter what $Q_1$ happens to be.
By contrast, values of $q_n$
may be chosen so that each eigenmode of the system with non-trivial radial structure
(i.e. $m>0$) may be stably influenced by the shear $Q_1$,  yet, \emph{there is no such configuration
of the disturbed shear flow that can act to stabilize a channel mode.}
This result, above all others, emphasizes the important uniqueness of these modes.
Also worth noting is that while the corrections to the growth rates
of all modes (except $m=0$) is proportional to $\epsilon$,
the corresponding growth rate correction for the channel mode scales as
 $\epsilon^2$: which means that though the channel mode is consistently destabilized
 by any periodic shear profile, its influence is markedly weaker.

\subsubsection{Periodic Conditions}\label{small_Q_periodic_conditions}
  {Solutions of $\psi_{0m}$ that are periodic on the domain $L$
are given by}
\beq
 \psi_{0m} =  A_m \sin \Bigl[\beta_m(x+L/2)
+\phi_m\Bigr],
\label{Periodic_small_Q_theory_psi_solution}
\eeq
  {where $ \beta_m \equiv {2m \pi}/{L}$ and $\phi_m$ is an arbitrary phase.}
The temporal response is given by $\sigma_{{0m}}^2(\beta_m^2)$,
as defined in (\ref{general_temporal_response_classic_MRI}).  As in
the previous section, if attention is first given to those modes where $m\neq 0$,
then one finds that
\beq
\sigma_{1m}^2 = q_{2m}\sfrac{1}{2}(-1)^m\sin\left(\frac{\pi}{2} - 2\phi_m\right)\Upsilon(m).
\label{sigma_1_periodic_bc}
\eeq
As in the previous section, the correction $\sigma_{1m}^2$ for the channel
mode ($m=0$) is zero and the first non-trivial temporal response comes at
order $\epsilon^2$.  The procedure is exactly the same as found in Section 5.2.2 and
its correction $\sigma^2_{20}$ is
\beq
\left(\frac{\partial\kappa_0^2}{\partial\sigma^2}\right)_{\sigma^2_{00}} \sigma^2_{20}
= \left(\frac{2\Omega_0^2 k^2}{\sigma_{00}^2 + \omega_{az}^2}\right)^2< (\partial_x\tilde{\cal Q})^2> ,
\eeq
with
\beqa
& & \tilde {\cal Q} = - \sum_{n=1}^{\infty}
q_n \left(\frac{L}{2n\pi}\right)^2\Biggl[
\sin\left(\frac{2n\pi}{L}x + \frac{\pi}{2}\right)\Biggr] .
\eeqa
Imposing periodic boundary conditions also implies that there are values of the
coefficients of the perturbed shear $Q_1$ which can be chosen to promote
stability for all the modes of the system except for the channel mode.
For example, for the $m=1$ mode and given its relative phase $\phi_1$
one can choose $q_2$  so that the RHS of (\ref{sigma_1_periodic_bc})
is negative.  This can be done for all other modes of the system as well except
for the $m=0$ channel mode, which is destabilized no matter
what choice is made for the parameters of $Q_1$, i.e.  the set $q_n$.
\subsection{Single shear defect on an infinite domain:
localized disturbances}\label{single_defect}
A profile with a single shear defect located at $x=0$ is represented
by the delta-function profile
\beq
Q = Q_0L \delta(x). \label{defect_Q}
\eeq
The corresponding departures from the mean Keplerian shear profile $U$
is given by
\beq
U =
 \left\{
\begin{array}{rl}
-Q_0 \Omega_0 L/2 , \ \ \  & x<0; \\
Q_0 \Omega_0L/2, \ \ \ & x>0.
\end{array}
\right.
\eeq
If $\psi_{_\pm}$ represents the solution of (\ref{MHD_psi_eqn})
respectively for $x>0$ and $x<0$, then solutions which decay as
$x \rightarrow \pm \infty$ are
\beq
\psi_{_\pm} = A_{\pm} e^{\mp |\kappa_0 |x}, \label{psi_sol_infinite_domain_defect}
\eeq
where $\kappa_0$ is as given in (\ref{kappa_0_def}) \emph{and provided that} $\kappa_0^2 > 0$.
If the latter condition is not met, then there are no localized normal modes allowed.
  { The coefficients
of the streamfunctions in both regions are equal in order that they
be continuous across $x=0$, thus $A_+ = A_- = A$.}
Integrating (\ref{MHD_psi_eqn}) in an infinitesimal region around
$x=0$, and given the defect (\ref{defect_Q}), imposes a jump condition
for the derivatives of $\psi$ approaching either side of the defect,
\beq
\partial_x\psi\bigr|_{0^+} - \partial_x\psi\bigr|_{0^-}
=
-\frac{2\Omega_0^2 k^2}{\sigma^2 + \omega_{az}^2} Q_0 L A.
\label{jump_condition}
\eeq
Putting in the solutions for the respective domains
gives the quantization criterion
\beq
|\kappa_0| = \frac{\Omega_0^2 k^2}{\sigma^2 + \omega_{az}^2} Q_0 L.
\label{defect_quantization}
\eeq
Solutions of the above equation are constrained to two possibilities:
(i) if $Q_0 < 0$, then $\sigma^2 + \omega_{az}^2 < 0$ for a normal mode solution
to exist while (ii) if $Q_0 > 0$, then it must be that $\sigma^2 + \omega_{az}^2 > 0$
for there to be a normal mode. The first of these cases says that normal modes are allowed only
for those values of $\sigma^2$ which are negative which are associated
with the HI modes.  The second of these cases are associated with the HMI
modes of which the MRI is a possibility.  Thus it is an interesting result
that the MRI is a permitted localized normal mode disturbance only if the shear defect is positive.
If the shear defect is negative (so that there is an effective reduction of
the shear inside the region), then there are no normal modes predicted.\par
The solution of (\ref{defect_quantization}) is obtained by first taking the square
of both sides of (\ref{defect_quantization})
yielding,
\beq
1-\frac{
4\Omega_0^2 \omega_{az}^2}
{(\sigma^2 + \omega_{az}^2)^2}
+\frac{\omega_0^2}{\sigma^2 + \omega_{az}^2}
=  \frac{\Omega_0^4 k^2}{(\sigma^2 + \omega_{az}^2)^2} Q_0^2 L^2. \label{longer_condition}
\eeq
  {However one must be careful in interpreting the solutions of this simplified
equation. Naively solving (\ref{longer_condition})
will yield twice as many solutions than are allowed.  This overcounting
is corrected by requiring that only those solutions of (\ref{longer_condition}) that
satisfy
\[
{Q_0}/({\sigma^2 + \omega_{az}^2}) >0,
\]
are permitted
or else (\ref{defect_quantization}) cannot be satisfied.}  Solutions
of (\ref{defect_quantization}) are,
\[
\sigma^2 + \omega_{az}^2 =
\sfrac{1}{2}
\left[
-\omega_0^2 \pm \sqrt{
\omega_0^4 + 4(Q_0^2 k^2 L^2 \Omega_0^4 + 4\Omega_0^2 \omega_{az}^2)}
\right].
\]
  {Because the term inside the radical sign is always greater than
zero for $q_0>0$, the ``+" branch of the above expression is always greater than zero
while the ``-" branch is always less than zero.  Thus the ``+" branch
is the permitted solution for $Q_0>0$ while the ``-" branch is the
allowed one for $Q_0<0$.   Incorporating this formally yields the
expression}
\[
\sigma^2 + \omega_{az}^2 =
\sfrac{1}{2}
\left[
-\omega_0^2 + {\rm sgn}(Q_0)\sqrt{
\omega_0^4 + 4(Q_0^2 k^2 L^2 \Omega_0^4 + 4\Omega_0^2 \omega_{az}^2)}
\right].
\]
  {Rewriting the terms appearing in the interior of the radical sign
leads to finally,}
\beqa
 \sigma^2 &=& -\left(\omega_{az}^2 + \frac{\omega_0^2}{2}\right)
 + {\rm sgn}(Q_0)\Biggl[
\left(\omega_{az}^2 + \frac{\omega_0^2}{2}\right)^2 \nonumber \\
& & \ \ \ \
-\omega_{az}^2\left(\omega_{az}^2 - {2\Omega_0^2 q_0 }\right) + \Omega_0^4 (kL)^2 Q_0^2
\Biggr]^{1/2},
\label{defect_sigma_sol}
\eeqa
which is similar in content to the classical channel mode dispersion relation
(\ref{sigma_0_sol}) except for the fact that at any one time there exists only
one branch of modes as a possible normal mode solution.   The MRI mode only
manifests as a normal mode when $Q_0 > 0$ and its growth rate is
enhanced by the defect.
\subsection{Symmetric shear step on an infinite domain: localized disturbances}\label{symmetric_shear_step}
A finite version of the shear profile evaluated in the previous section is,
\beq
Q =
 \left\{
\begin{array}{rl}
0, \ \ \ &  x < -\frac{L}{2}; \\
Q_0 , \ \ \  & -\frac{L}{2}< x< \frac{L}{2}; \\
0, \ \ \ &  x > \frac{L}{2};
\end{array}
\right.
\label{finite_shear_step_profile}
\eeq
This profile for $Q$ has a \emph{top-hat} structure and this descriptor will be used
interchangeably with the expression \emph{symmetric shear step}.
Evidentally the value of the integral, $\int^{\infty}_{-\infty} Q dx = Q_0 L$, is the
same for both forms of $Q$ given in (\ref{defect_Q}) and (\ref{finite_shear_step_profile}).
Solutions to (\ref{MHD_psi_eqn}) must be developed separately in the three
regions and appropriately matched across the boundaries separating the regions.
  {Because $Q$ is bounded} it will suffice to match both
the streamfunctions and their first derivatives across the domain.
Thus, for $x<-L/2$ and $x>L/2$ (\ref{MHD_psi_eqn}) is
\beq
\partial_x^2 \psi - \kappa_0^2 \psi = 0.
\eeq
The solutions in the region where $x<-L/2$ is given by
\beq
\psi_{_-} = A_- e^{|\kappa_0|(x+L/2)},
\eeq
while for $x>L/2$
\beq
\psi_{_-} = A_+ e^{-|\kappa_0|(x-L/2)},
\eeq
together with the constraint that $\kappa_0^2 > 0$.  In the region
$-L/2 < x < L/2$ (\ref{MHD_psi_eqn}) is
\beq
\partial_x^2 \psi - \left(\kappa_0^2 -\frac{2\Omega_0^2 Q_0 k^2}{\sigma^2 + \omega_{az}^2}\right) \psi = 0,
\eeq
which has the two linearly independent solutions
\footnote{The second of the solutions
appearing in (\ref{stepQ_psi_sol_interior_zone})
 is written in the form displayed in order to make
sure that both linearly independent solutions are represented even in the event
$\kappa =0$ (Friedman 1956).}
\beq
\psi_0 = A_0 \cosh[\kappa x] + B_0 \frac{\sinh[\kappa x]}{\kappa}.
\label{stepQ_psi_sol_interior_zone}
\eeq
in which
\beq
\kappa^2(\kappa_0) \equiv \kappa_0^2-\frac{2\Omega_0^2 Q_0 k^2}
{\Sigma^{(\pm)}(\kappa_0,\omega_0)}. \label{kappa_squared_def}
\eeq
The functional form
\beq
\sigma^2 + \omega_{az}^2 = \Sigma^{(\pm)}(\kappa_0,\omega_0)
\label{general_sigma_solution}
\eeq
in which
\beqa
& & \Sigma^{(\pm)}\left(\kappa_0,\omega_0^2\right)\equiv \nonumber \\
& &  \frac{1}{2\left(1-\frac{\kappa_{0}^2}{k^2}\right)}
\left[-\omega_0^2 \pm
\left[\omega_0^4 + 16\Omega_0^2 \omega_{az}^2\left(1-\frac{\kappa_{0}^2}{k^2}\right)\right]^{1/2}
\right].
\eeqa
is introduced in order to make the following discussion more transparent.
The $\pm$ designation references the HMI or HI modes respectively.
It is also worth noting that given the constraint that $\sigma^2$ must be
real, that there is an absolute maximum value allowed for $\kappa_0^2$ and
this is dictated by the requirement that the terms subject to the square root
operator must remain greater than zero or else the value of $\sigma^2$
will turn out complex.    {In other words, given the reality of $\sigma^2$
one does not expect solutions of
$\kappa_0^2 $  to exceed $ \kappa_{max}^2$ where}
\beq
\kappa_{max}^2
= k^2\left(\frac{\omega_0^4}{16\Omega_0^2\omega_{az}^2} + 1\right).
\eeq
  {In all of the numerical solutions calculated below it is found that, indeed, all the
solutions determined satisfy $\kappa_0^2 < \kappa_{max}^2$.}
 \par
Imposing the conditions that the streamfunction and its first derivative match
across the two boundaries $x=\pm L/2$ reveals that normal mode solutions exist
provided the following condition is satisfied
\beq
\left[
\tanh\left(\kappa \frac{L}{2}\right) +\frac{\kappa}{|\kappa_0|}\right]
\left[
\coth\left(\kappa \frac{L}{2}\right) +\frac{\kappa}{|\kappa_0|}\right] = 0.
\eeq
The terms in the first bracket represent a quantization condition for odd-parity
modes, which is to say that in the dimpled region $\psi_0 \sim \sinh [\kappa x]/\kappa$,
 while the second represents even-parity modes where similarly
 $\psi_0 \sim\cosh[\kappa x]$ in the same region.\par
Attention is first given
to the even-parity mode in which the quantization condition,
\[
\coth\left(\kappa \frac{L}{2}\right) = -\frac{\kappa}{|\kappa_0|},
\]
must be satisfied for some value of $\kappa_0$.
If $\kappa$ were real and greater than zero, then the LHS of the above relationship is always
positive.     {Since $\kappa_0^2$ is real and positive}, if the modes are
to be localized, then the RHS of this expression
is always less than zero. In this case it means that
no solution exists for $\kappa$ real.
On the other hand solutions do exist if
$\kappa^2 \le 0$.  Rewriting
$\kappa = i\tilde \kappa$ then the quantization condition above becomes
\beq
\cot[\tilde \kappa L/2]  = \frac{\tilde\kappa}{|\kappa_0|},
\label{quantization_condition_odd}
\eeq
now with the constraint for the existence of localized normal modes
is re-expressed as
\beq
\tilde\kappa^2(\kappa_0) \equiv \frac{2\Omega_0^2 Q_0 k^2}{\Sigma^{(\pm)}(\kappa_0,\omega_0)}
-\kappa_0^2 > 0.
\label{constraint_1}
\eeq
The strategy for obtaining a solution
is as follows: find the values of $\kappa_0$ which
solve (\ref{quantization_condition_odd}),
subject to the constraint given in (\ref{constraint_1}),
and then given the solution $\kappa_0$ find
 $\sigma^2$ through the relationship
(\ref{general_sigma_solution}).  The same procedure will
be implemented for the even-parity solution below as well.
Solutions of the quantization
condition will be sought graphically in the range of values
for $\kappa_0 \in (0,\kappa_{max})$.  Inspection shows that
the constraint (\ref{constraint_1}) is violated
if $Q_0 < 0$ and $\Sigma^{(\pm)}>0$,
or if $Q_0 > 0$ and $\Sigma^{(\pm)}< 0$.
Since $\Sigma^{(+)}>0$ it means that there are no HMI normal modes
as allowable solutions if $Q_0 < 0$.  HI normal modes are associated
with the $\Sigma^{(-)}$ branch of solutions and since this is less
than zero \emph{except for very special circumstances}, one can
generally expect HI normal modes for $Q_0 < 0$ and for them to be stable.
The aforementioned special circumstance occurs
if there are solutions to
(\ref{quantization_condition_odd}) in which $\kappa_0 > k$ since,
in that case,  $\Sigma^{(-)} > 0$ and HI modes may exist as well for $Q_0 > 0$.
\par
\begin{figure}
\begin{center}
\leavevmode \epsfysize=8.6cm
\epsfbox{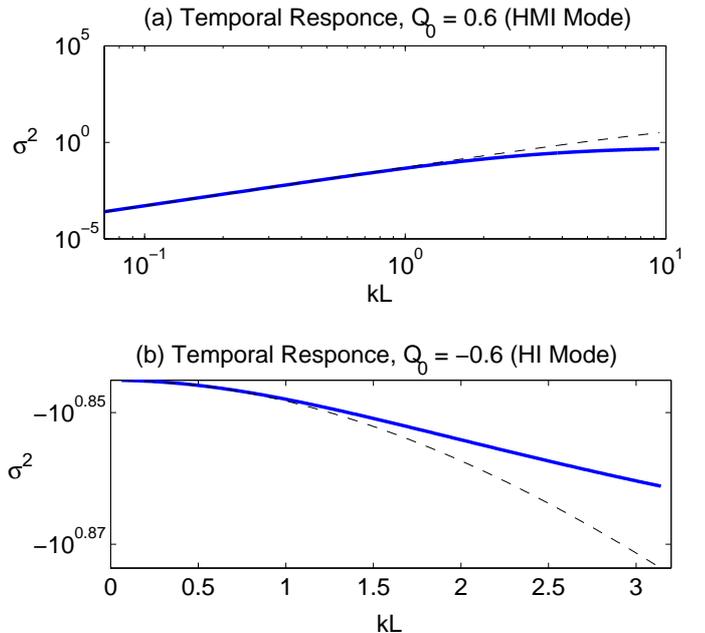}
\end{center}
\caption{The temporal response resulting from the quantization condition
(\ref{quantization_condition_odd}) as a function of
$kL$ (solid line) compared to the quantization condition (\ref{defect_quantization})
for defect profile (dashed line).  For both plots $\omega_{az}^2 = 2\Omega_0^2 q_0$
with $q_0 = 3/2$.  (a) For $Q_0 > 0$ only HMI normal modes are permitted.
The agreement between the two profiles severely breaks down for $kL$ exceeding $2$.
(b) For $Q_0 < 0$ only HI normal modes are allowed.  The agreement between
defect theory and step profile is better for HI modes.
}
\label{section5p4plot1}
\end{figure}

The even-parity mode limits to the single-defect configuration discussed in
Section \ref{single_defect} for small values of the
parameter $L$.  Analysis of
(\ref{quantization_condition_odd}) in this limit shows that,
\[
\frac{2}{\tilde \kappa L} \approx \frac{\tilde\kappa}{|\kappa_0|},
\]
which, after restoring the definition of $\tilde\kappa$ and some rearranging, becomes
\[
|\kappa_0| = \frac{L}{2}\left(
\frac{2\Omega_0^2 Q_0 k^2}{\sigma^2 + \omega_{az}^2}
-\kappa_0^2\right),
\]
where the definition of $\Sigma^{(\pm)}$ has been restored.  For values of
$Q_0$ such that $\Omega_0^2Q_0 \gg (\sigma^2 + \omega_{az}^2)\kappa_0^2/k^2$
the above expression reduces to
\[
|\kappa_0| =
\frac{\Omega_0^2 Q_0 L k^2}{\sigma^2 + \omega_{az}^2},
\]
which is the quantization condition for the single-defect problem on
the infinite domain, (\ref{defect_quantization}).
The range of values for $Q_0$ and $L$ for which this limiting
form agrees with the actual result of the quantization condition
is shown in Figure \ref{section5p4plot1}, where $\omega_{az}^2$ is
set to the value $2\Omega_0^2 q_0$ in order to compare
the temporal response against conditions in which the channel mode
is marginal in the classical theory, see (\ref{velikhov_condition}).
The main result is that the temporal response of the defect profile
of the previous section closely resembles the temporal response
of this step profile here in the limit where $kL$ is small less than
one.  The limiting form represented by the defect profile becomes
a poor representation of the top-hat profile when $kL$ exceeds order 1
values, however, the agreement tends to be much better for the HI.
\par
\begin{figure}
\begin{center}
\leavevmode \epsfysize=6.2cm
\epsfbox{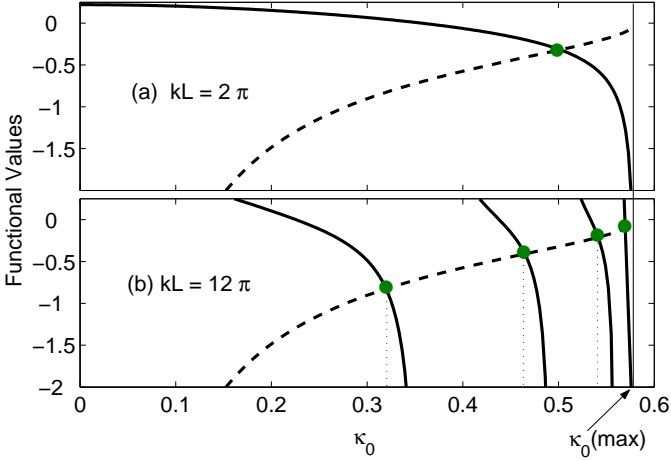}
\end{center}
\caption{Plot showing the increase of allowed normal modes with
domain size $L$.  In all plots $\omega_{az}^2 = 2\Omega_0^2 q_0$,
$q_0 = 3/2$ and $Q_0 = 0.6$ (HMI mode permitted).  The dashed
line is $\tilde\kappa/|\kappa_0|$ and solid line is $\cot \tilde\kappa L/2$.
Panel (a): $kL = 2\pi$ corresponds to one normal mode
at $\kappa_0 =0.50$. Panel (b): $kL = 12\pi$ supporting
four normal modes $\kappa_0 = 0.32, 0.465, 0.54, 0.56$.  All modes correspond
to values of $\sigma^2 > 0$.
$\kappa_{0}({\rm{max}}) = 0.576$.
}
\label{section5p4plot2}
\end{figure}

Additionally, for small values of $kL$ there is always at least one normal mode
expected, whether it be the HI or HMI mode.  However,
as the horizontal length of the symmetric step profile increases, the number
of permitted normal modes increases as well (Figure \ref{section5p4plot2}).
Similar behavior is predicted for modes which are localized in
the vertical direction of the disk (Liverts \& Mond 2009).
As the number of permitted normal modes increases
their $\kappa_0$ values will cluster (countably as $kL \rightarrow \infty$)
around the maximum value
\beqa
& & \Bigl(\kappa_0^2\Bigl)_{max} = \nonumber \\
& & \ \ \frac{Q_0}{4\omega_{az}^2}
\left(\omega_0^2 - 2Q_0\Omega_0^2
+\sqrt{(\omega_0^2 - 2Q_0\Omega_0^2)^2 + 16\Omega_0^2\omega_{az}^2}
\right).
\eeqa
\par
Odd-parity modes are interesting since they have no analogues in
the single defect profile studied in the previous section.
Examining the quantization condition for it,
\[
\tanh\left(\kappa \frac{L}{2}\right) +\frac{\kappa}{|\kappa_0|} = 0,
\]
shows that there are no real solutions of $\kappa$ that satisfy the constraint.
Utilizing the expression $\kappa = i\tilde\kappa$, as used previously, the
quantization condition takes the form
\beq
\tan[\tilde \kappa L/2]  = -\frac{\tilde\kappa}{|\kappa_0|}.
\label{quantization_condition_even}
\eeq
Inspection of the condition (\ref{quantization_condition_even}) shows
that for there to exist a normal mode $\tilde\kappa$ must be greater
than $\pi/L$ as the functional form
$\tan[\tilde \kappa L/2]/\tilde\kappa$ is positive for values
of $\tilde\kappa \in (0,\pi/L)$.  This places a constraint upon the
minimum value of $L$ that permits this mode to exist and this obtained
by solving
\beq
\tilde\kappa = \frac{\pi}{kL},
\eeq
and, using the definition for $\kappa$ found in (\ref{kappa_squared_def})
one finds that  the minimum value
for $L$ must satisfy,
\beq
kL_{min} = \left[\frac{\omega_0^2\pi^2}{4 Q_0}\left({\rm sgn}(Q_0)
\sqrt{1+\frac{16\Omega_0^2\omega_{az}^2}{\omega_0^4}}-1\right)
\right]^{1/2}. \label{min_L_criterion_odd_parity_mode}
\eeq
Note that the inclusion of the expression
${\rm sgn}(Q_0)$ reflects how HMI/HI modes
associate with the sign of $Q_0$ so that the expression
appearing above encapsulates both possibilities.
Odd-parity modes bifurcate into existence for $L\rightarrow\infty$
as $Q_0$ grows from zero and therefore it means that these
modes are not expected to play a role for values of the defect
which are small as this would entail very large values of the
top-hat region.
The temporal response of these odd-parity modes are otherwise similar
in most every respect to their even-parity counterparts and,
as such, no further analysis of their properties is pursued
here.

It should be noted that as in the previous section unstable
localized MRI normal modes (HMI-modes) can exist only
if the step function in the shear is greater than zero.  If the step is negative
(indicating a region of weakened shear), then there are no HMI types of localized normal modes
admitted by the configuration and, hence, there is no possibility
of normal mode unstable disturbances.\par
\subsection{Single shear defect on a periodic domain}\label{ShearDefectPeriodicDomain}
Next a shear defect of the form
\beq
Q = -Q_0 + Q_0 L \delta(x),
\eeq
is considered on the periodic domain $-L/2\le x \le L/2$.
This form is very
similar to the one considered in Section \ref{single_defect} except
the integral of the shear over the periodic domain is zero, i.e.
$<Q> = 0$.
The effects of shear defects with zero mean may be examined.
The governing equation for $\psi$ is
\beq
\partial_x^2 \psi - \kappa_{_{00}}^2\psi = -\frac{2\Omega_0^2k^2 Q_0 L \delta(x)}{\sigma^2 + \omega_{az}^2},
\label{periodic_defect_psi_eqn}
\eeq
where  $\kappa_{_{00}}$ is given by
\beq
\kappa_{_{00}}^2 =
k^2\left(1 + \frac{4\Omega_0^2 \sigma^2}{(\sigma^2 + \omega_{az}^2)^2}
- \frac{2\Omega_0^2(q_0-Q_0)}{\sigma^2 + \omega_{az}^2}
\right).
\eeq
Similar to what was done in the previous sections, one may
express $\sigma$ in terms of $\kappa_{_{00}}$ via the expression $\Sigma^{(\pm)}$
and, for the sake of completeness, this is written out explicitly,
\beqa
& & \sigma^2 + \omega_{az}^2 = \Sigma^{(\pm)}\left(\kappa_{_{00}},\omega_{_{00}}^2\right)\equiv \nonumber \\
& & \ \frac{1}{2\left(1-\frac{\kappa_{_{00}}^2}{k^2}\right)}
\left[-\omega_{_{00}}^2 \pm
\left[\omega_{_{00}}^4 +
16\Omega_0^2 \omega_{az}^2\left(1-\frac{\kappa_{_{00}}^2}{k^2}\right)\right]^{1/2}
\right],
\label{Sigma_2_definition}
\eeqa
in which the shear modified epicyclic frequency is defined as
$\omega_{_{00}}^2 \equiv 2\Omega_0^2(2-q_0+Q_0)$.
Solutions
to (\ref{periodic_defect_psi_eqn}) must be developed separately for either side of
$x=0$ and matched.  Solutions for the streamfunction
in which $\psi(-L/2) = \psi(L/2)$
and $\partial_x\psi|_{-L/2} = \partial_x\psi|_{L/2}$ (i.e. that they are periodic)
are
given by
\beq
\psi_{_\pm} = A\cosh\left[\kappa_{_{00}}\left(x\mp \frac{L}{2}\right)\right],
\label{periodic_defect_psi_sol}
\eeq
where $\psi_{_\pm}$ is the solution for the region $x\in (0,\pm L/2)$.
Continuity of the stream function is ensured by
construction of the form of the solution in (\ref{periodic_defect_psi_sol}).  The
streamfunction must show a jump in its first derivative following the same
arguments found in Section \ref{single_defect}.  Integrating (\ref{periodic_defect_psi_eqn})
in a small region around $x=0$ results in the same condition for the jump in $\partial_x\psi$
found in (\ref{jump_condition}).  Putting in the forms for $\psi_{_\pm}$ found
in (\ref{periodic_defect_psi_sol}) leads to the quantization condition
\beq
\kappa_{_{00}} \tanh\left[\kappa_{_{00}}\frac{L}{2}\right]
= \frac{\Omega_0^2k^2 Q_0 L}{\Sigma^{(\pm)}\left(\kappa_{_{00}},\omega_{_{00}}^2\right)}
={\cal F}^{(\pm)}\left(\kappa_{_{00}},\omega_{_{00}}^2\right) .
\label{periodic_defect_quantization}
\eeq
It should be kept in mind that the
quantities ${\cal F}^{(\pm)}$ as appearing on the RHS of (\ref{periodic_defect_quantization})
are different functions of $\kappa_{_{00}}$ depending upon whether one is
considering HMI/HI ($\pm$) modes, respectively.  Thus when solutions
of the quantization condition are sought, separate attention will be
given depending upon which type of mode is of interest.
Furthermore,
consideration of the solutions to (\ref{periodic_defect_quantization}) must be
done by restricting attention to cases where $\kappa_{00}$ is either real
or imaginary. Additionally from previous arguments showing that $\sigma^2$ must
be real, there will be a maximum real value possible for $\kappa_{00}$
\beq
\kappa_{max}^2\left(\omega_{_{00}}^2\right)
= k^2\left(\frac{\omega_{_{00}}^4}{16\Omega_0^2\omega_{az}^2} + 1\right).
\eeq
Values of $\kappa_{_{00}}$ which are greater that $\kappa_{max}\left(\omega_{_{00}}^2\right)$
would lead to complex values of $\sigma^2$ as an inspection of
(\ref{Sigma_2_definition}) shows that if $\kappa_{00}$ does exceed this
maximum value, then the term inside the square-root operator becomes negative
which would lead to a complex value of $\sigma^2$.  This restriction upon
the value of $\kappa_{_{00}}$ helps in finding solutions for it.
\par
The limit where $Q_0 \rightarrow 0$ recovers the classical limits discussed
in Section \ref{classical_limit_theory}.  In that case channel
modes, i.e. those modes for which $\kappa_0 = 0$, are the most unstable.
In the current setting
this corresponds to $\kappa_{_{00}} = 0$ when $Q_0$ = 0.  Additionally,
for $\omega_{az}^2 = 2\Omega_0^2 q_0$ the channel mode is exactly
marginal for $Q_0 = 0$, viz. the arguments leading to (\ref{velikhov_condition}).
This value for  $\omega_{az}^2$ will be
assumed so that a controlled analysis may be carried out in which
one can track how varying $Q_0$ affects the stability characteristics
of an otherwise marginal channel mode.
\par
\begin{figure}
\begin{center}
\leavevmode \epsfysize=9.2cm
\epsfbox{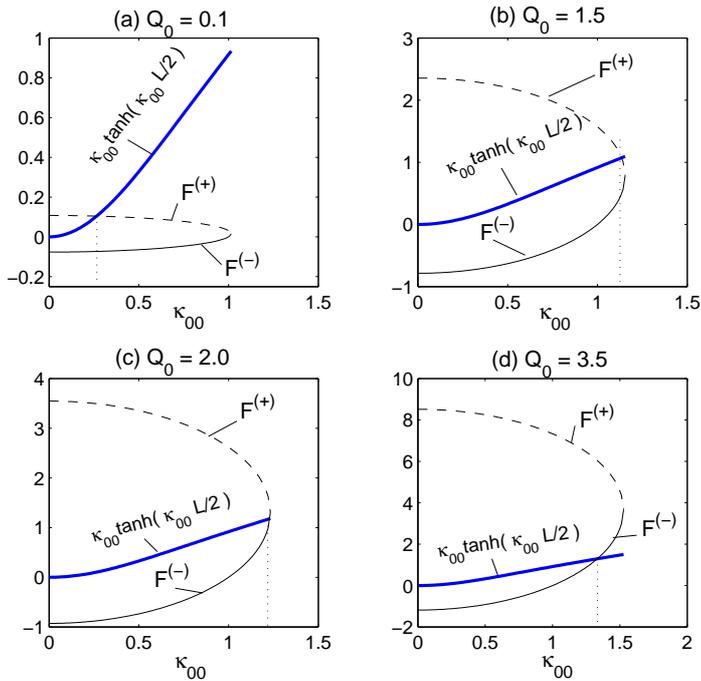}
\end{center}
\caption{Graphical solutions for the quantization condition (\ref{periodic_defect_quantization}).
Attention restricted to real values of $\kappa_{_{00}}$ marking the mode
which becomes the classical channel mode in the limit $Q_0 \rightarrow 0$.
Each panel depicts
the three functions $\kappa_{_{00}}\tanh\kappa_{_{00}}L/2$ (thick line), and ${\cal F}^{(\pm)}$
(dashed/thin lines).
In all plots $kL = \pi$ and $\omega_{az}^2 = 2\Omega_0^2 q_0$.
Panels (a) and (b) show that HMI modes are admitted
as the solid curve crosses the ${\cal F}^{(+)}$ line while for the latter two
it crosses the line corresponding to ${\cal F}^{(-)}$ meaning that HI modes
are selected.  The solution $\kappa_{_{00}}$ and temporal response for each:
(a) $Q_0 = 0.1, \kappa_{_{00}} = 0.27,  \sigma^2 = 0.005$
,(b) $Q_0 = 0.5, \kappa_{_{00}} = 0.65, \sigma^2 = 0.14$,
(c) $Q_0 = 2.0, \kappa_{_{00}} = 1.33,\sigma^2 = 3.36$ and
(d) $Q_0 = 3.5, \kappa_{_{00}} = 1.37, \sigma^2 = 6.73$.  Notice how the channel mode goes from
being an unstable HMI mode to an unstable HI mode
as $Q_0$ is increased.
}
\label{section5p5plot1}
\end{figure}
\begin{figure}
\begin{center}
\leavevmode \epsfysize=9.cm
\epsfbox{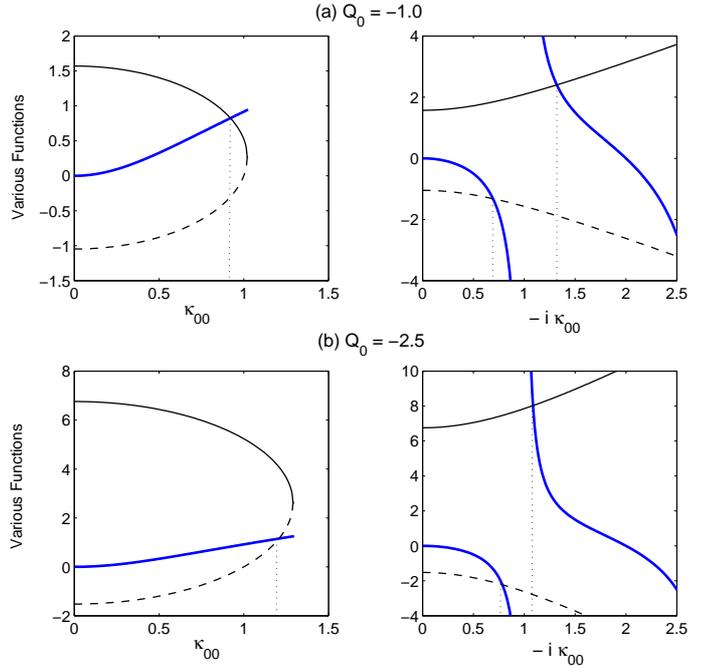}
\end{center}
\caption{
Similar to Figure \ref{section5p5plot1} except negative values of $Q_0$
are investigated. In all plots $kL = \pi$ and $\omega_{az}^2 = 2\Omega_0^2 q_0$
and $q_0 = 3/2$.
Panel (a): the left plot shows that the channel mode admitted is a stable HI mode
with $\kappa_{_{00}} =0.92$ and $\sigma^2 = -5.30$.  The right plot in Panel (a)
shows the first two overtone modes.  The first HMI mode is unstable
with $\kappa_{_{00}} = 0.69 i$ and $\sigma^2 = 0.88$ while the next HI
mode admitted has $\kappa_{_{00}} = 1.32 i$ and $\sigma^2 = -2.88$.  Note that in these
cases $\kappa_{_{00}}$ is imaginary.
Panel (b):  stable HMI mode
with $\kappa_{_{00}} =1.21$ and $\sigma^2 = -8.30$. For the
first two overtones,  the HMI mode is unstable
with $\kappa_{_{00}} =0.78 i$ and $\sigma^2 = 2.02$ while for the HI
mode, $\kappa_{_{00}} =1.08 i$ and $\sigma^2 = -2.47$.  For all values
of $Q_0 < 0$ there is at least one unstable HMI mode with imaginary $\kappa_{_{00}}$.
}
\label{section5p5plot2}
\end{figure}
\begin{figure}
\begin{center}
\leavevmode \epsfysize=8.5cm
\epsfbox{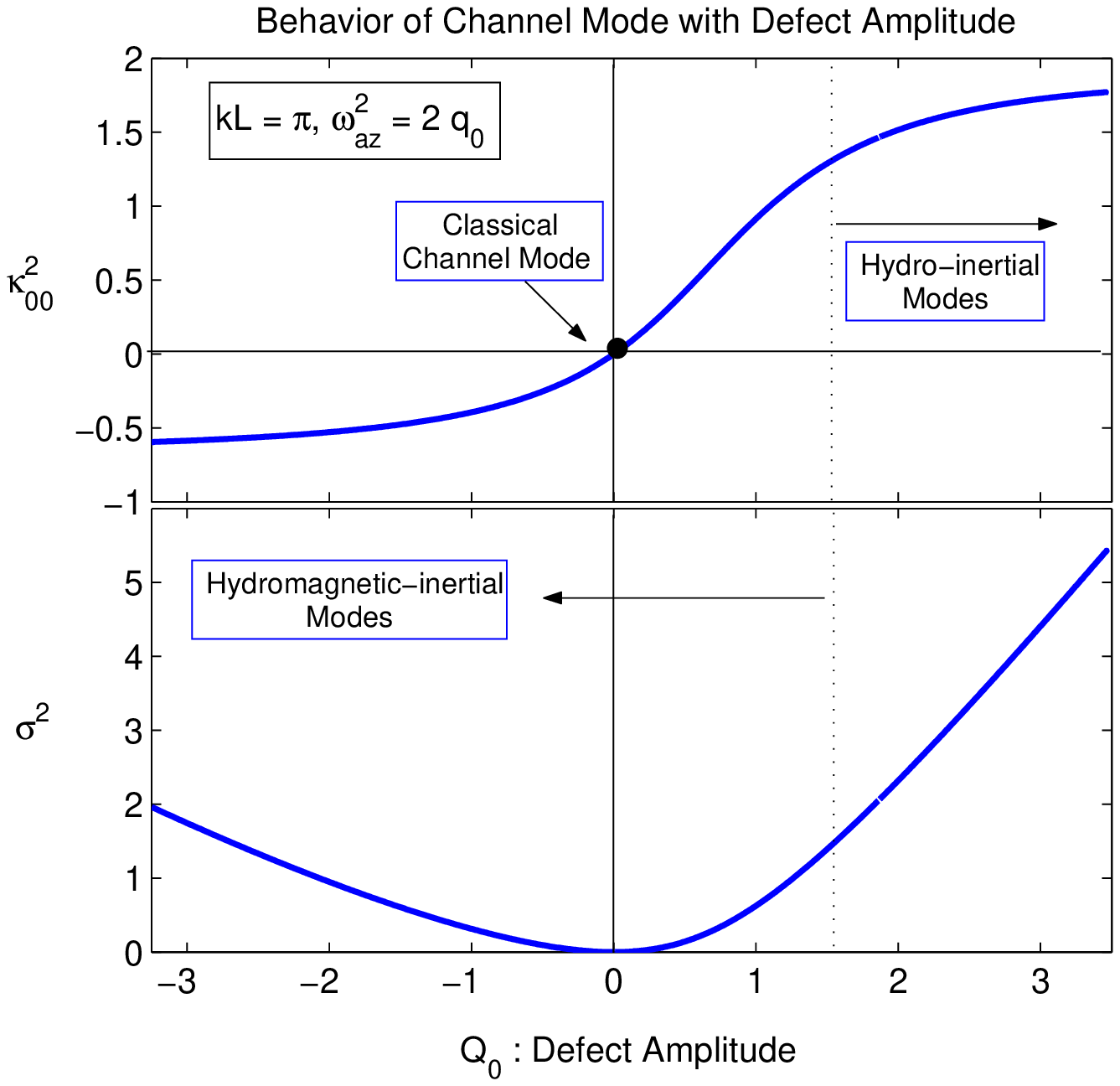}
\end{center}
\caption{
Behavior of the classical channel mode as a function of defect amplitude $Q_0$
for $kL = \pi$ and $\omega_{az}^2 = 2\Omega_0^2 q_0$, $q_0 = 3/2$, corresponding to a marginal
($\sigma^2 =0$)
channel mode at $Q_0 = 0$.  Top panel depicts $\kappa_{_{00}}^2$ as a function
of $Q_0$ in which the classic channel mode has $\kappa_{_{00}}^2 = 0$.  Bottom
panel shows the temporal response which predicts exponential growth/decay for
all values $Q_0 \neq 0$.  The channel mode goes from being a HMI mode to
a HI mode at $Q_0 \approx 1.87$.
}
\label{section5p5plot3}
\end{figure}

Figures \ref{section5p5plot1}-\ref{section5p5plot2} show the graphical solutions
of (\ref{periodic_defect_quantization}) with some details of the results.  The main
result is that the mode that is identified with the classical channel mode
in the ideal limit always exhibits exponential temporal growth
no matter what amplitude the shear defect
takes.    Figure \ref{section5p5plot3}
shows that the temporal response of the classical channel mode is maximal for
$Q_0 = 0$ and that as $Q_0$ moves away from zero instability is predicted.
The mode exhibits an exponentially decaying spatial character
($\kappa_{_{00}}^2 > 0$) for
$Q_0 > 0$, while it is weakly oscillatory ($\kappa_{_{00}}^2 <0$) for
$Q_0 < 0$.  This general trend persists for a wide range
of $kL$ values. As such, the results depicted in these figures
can be taken to be qualitatively representative.
The figures also indicate that the unstable HMI mode can
also turn into an unstable HI mode
if the amplitude of $Q_0$ becomes large enough.

\subsection{Single Shear Defect on Finite Domain: Channel Boundary Conditions}\label{finite_domain_single_defect_channel_walls}
This section is concerned with the response of disturbances to a $Q$ configuration
that is exactly like that considered in Section \ref{ShearDefectPeriodicDomain},
where the difference
now is that the $\psi$ disturbances
are set to zero at the boundaries $x = \pm L/2$.  This
is equivalent to enforcing zero normal velocity at the boundaries.
Thence, $\psi$ in this case is
\beq
\psi_{_\pm} = \mp A\frac{\sinh\left[\kappa_{_{00}}\left(x\mp \frac{L}{2}\right)\right]}
{\sinh \kappa_{_{00}}L/2},
\label{channel_wall_defect_psi_sol}
\eeq
where just as in the previous section,
 $\psi_{_\pm}$ is the solution for the region $x\in (0,\pm L/2)$.  Evaluating the
 jump condition at $x=0$ leads to an analogous quantization condition
 like that in (\ref{periodic_defect_quantization}), namely,
\beq
\kappa_{_{00}} \coth\left[\kappa_{_{00}}\frac{L}{2}\right] =
\frac{\Omega_0^2k^2 Q_0 L }{\sigma^2 + \omega_{az}^2}
= \frac{\Omega_0^2k^2 Q_0 L}{\Sigma^{(\pm)}\left(\kappa_{_{00}},\omega_{_{00}}^2\right)},
\label{channel_wall_defect_quantization}
\eeq
in which ${\Sigma^{(\pm)}\left(\kappa_{_{00}},\omega_{_{00}}^2\right)}$ and
$\kappa_{_{00}}$ and $\omega_{_{00}}^2$ are exactly as they are found
in Section \ref{ShearDefectPeriodicDomain}.  It can be seen here, as before,
that there there are never real solutions of $\kappa_{_{00}}$ for HMI modes
if $Q_0 < 0$ and, likewise, no such solutions for HI modes
if $Q_0 > 0$.\par
\begin{figure}
\begin{center}
\leavevmode \epsfysize=8.5cm
\epsfbox{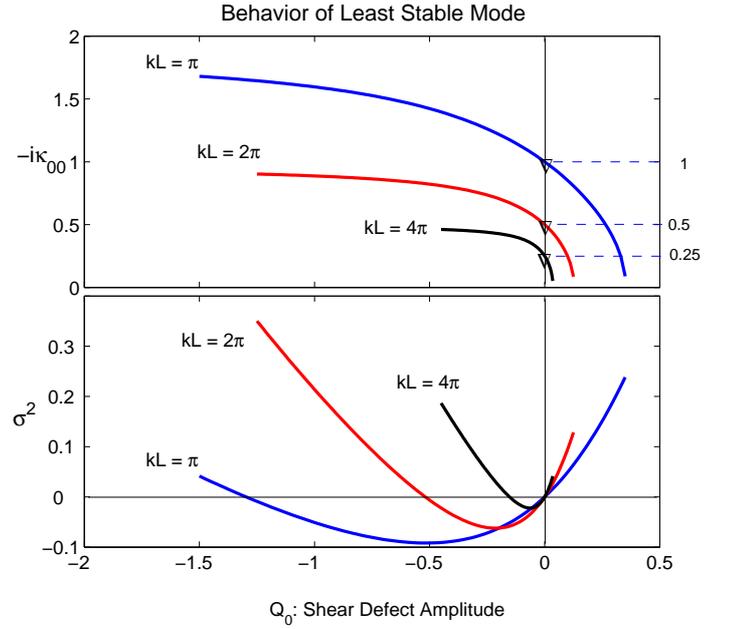}
\end{center}
\caption{The behavior of the first unstable (HMI) mode for the problem
with channel walls ($q_0 = 3/2$).  The values of $i \tilde\kappa_{_{00}} = \kappa_{_{00}}$
and the corresponding temporal response $\sigma^2$ are given as a function
of the defect amplitude $Q_0$ for three values of the parameter $kL$.  The top
panel shows $\tilde\kappa_{_{00}}$.
The bottom panel shows the temporal response.
The value of $\omega_{az}^2$ for each curve is set so that the temporal response
is marginal at $Q_0 = 0$ - details of this is described in the text.  The value
of  $\tilde\kappa_{_{00}}$ for $Q_0 = 0$ are shown by the $\nabla$ symbol in
the top panel.
}
\label{section5p6plot1}
\end{figure}

In Figure \ref{section5p6plot1} the stabilizing behavior of the shear defect
on the HMI modes
is depicted for the most unstable mode of the system.
The results from the weak shear limit discussed in Section
\ref{ChannelWalls_smallQ} are used as a reference point to
help interpret the following.
From the classical theory discussed in that section it
was established that the most unstable normal mode allowed
is the one in which $m=1$, in turn implying $\gamma_1 = \pi/L$.  Reference
to (\ref{criterion_for_instability_classic_mri}) shows that
this mode is marginally stable when
\beq
\omega_{az}^2 = 2\Omega_0^2 q_0 \frac{k^2L^2}{k^2L^2 + \pi^2}.
\label{omega_az_marginal_channel_wall}
\eeq
(This is obtained by replacing $\beta_n$ with $\gamma_1$ in that expression.)
 The response
of the modes
for three different values of $kL$
are depicted.  With the above choice for $\omega_{az}^2$
all modes are marginal at $Q_0 = 0$.  For the range of parameters
investigated and shown in the figure, $\kappa_{_{00}}$ must be imaginary
in order for the quantization condition to be satisfied.  Thus
the values for $\kappa_{_{00}}$ are written as $i\tilde\kappa_{_{00}}$.
Inspection of the results show that stability is promoted only for values
of $Q_0$ which are negative.  The range of values of $Q_0$ for
which stability is predicted decreases for larger domain
sizes, i.e. as $L$ is made larger the window in $Q_0$ for which
disturbances are stable decreases.  For those ranges in $Q_0$ in
which stability is predicted, e.g. for $Q_0 \in (-0.5, 0)$ for $kL = 2\pi$
with $\omega_{az}^2$ given in (\ref{omega_az_marginal_channel_wall}),
the temporal response of the other
overtones was checked and all were found to have $\sigma^2 < 0$.

\section{Discussion}
This study is devoted to examining the axisymmetric inviscid
linear response of a
shearing sheet environment threaded by a constant vertical
magnetic field for an array of different shear profiles.
The motivation for considering this setup
are restated from the Introduction:
\begin{enumerate}
\item  Numerical experiments of low magnetic Prandtl number
flows appropriate for cold disk environments
are outside current computational reach.
\item Current numerical experiments seem to show the
trend that as P$_{{\rm m}}$ is lowered so is
the corresponding angular momentum transport.  The reasons
for this is not yet clear.
\item Theoretical analysis of laboratory setups evaluating
the response of the MRI subject to a vertical background
field find that transport is reduced with P$_{{\rm m}}$.
Furthermore this reduction emerges through
the modification
of the basic shear profile (e.g. Taylor-Couette) inside the experimental cavity which,
in turn, is driven into place by the MRI itself.
The final condition is a stable pattern state.
\item In the final pattern state reached by the model laboratory system
the modified shear profile's amplitude is independent of
P$_{{\rm m}}$.  This is not true of the remaining fluid
and magnetic quantities which scale
as P$_{{\rm m}}^{\lambda}$ with typical values $\lambda \approx 1/2$.
As P$_{{\rm m}}$   {is made very small one would observe
a fluid state characterized by a uniform background magnetic field
and an azimuthal flow showing deviations from a Keplerian state.}
\item The main mode of instability and driver of turbulence
in the numerical experiments
of the shearing box is the channel mode which is a special response
with no radial structure.  This mode precipitates a secondary
instability which is understood to lead to a turbulent cascade.
There is no channel mode allowed in a setup appropriate for
a laboratory experiment.
\end{enumerate}
The saturation hypothesis is then the following: is it possible that
the fluid response in a shearing sheet environment
characterized by very low P$_{{\rm m}}$ stabilizes itself
by modifying the background shear
as suggested by the calculations appropriate for
laboratory conditions?  Of course,
there are some constraints placed upon the way in
which the shear may be modified in that the
gravitational field predominantly ``forces" there
to be a Keplerian profile (Ebrahimi et al. 2009).  However by analogy to the situation
for the laboratory setup, there may come into
play undulations of the shear whose average is zero over some
length scale $L$ (e.g. Figure 1).  \par
Of course, the main difference
between the laboratory setup and numerical
calculations of the shearing box are their respective boundary conditions.
The numerical experiments adopt periodic boundary conditions
while the calculations for the laboratory experiments require
no normal flow conditions at an inner and outer radius
and the former of these permits channel modes to come into play.
\par
Thus, what has been done here is a simple model calculation
in which a shearing box configuration
with a constant background vertical magnetic field is
tested for stability for a number of different shear
profiles.  The perspective taken is that if
there exists a shear profile (say, whose average over some
length scale is zero) that is stable to the MRI,
then it would strongly suggest that low P$_{{\rm m}}$ flow
configurations might saturate the MRI by driving
into place a new azimuthal flow profile which leads
to saturation.  \par
What is found here is something
very interesting.  For flow configurations in which
no-normal flow conditions are imposed at both
radial boundaries (Sections \ref{ChannelWalls_smallQ} \& \ref{finite_domain_single_defect_channel_walls})
there exists
a modification to the shear in which the most unstable
MRI mode is stabilized.
However, for flows in which periodic boundary conditions (Sections
\ref{small_Q_periodic_conditions} \& \ref{ShearDefectPeriodicDomain})
or fixed Lagrangian pressure conditions (Section \ref{small_Q_pressure_conditions})
are imposed there are no modulations of the shear
which shut off the instability.  In these cases,  all modifications
of the shear (with zero mean on some radial length scale) seem to enhance instability of the
channel mode.  For example, if the given channel
mode is marginally stable to exponential growth, then
introduction of the modified shear, no matter what is
its non-zero amplitude, seems to destabilize it as the results of
Section \ref{small_Q_periodic_conditions} and
Section \ref{ShearDefectPeriodicDomain} indicate.\par
This strongly implies that
in those numerical experiments in which a shear is threaded by a constant
vertical magnetic field,
the shear modification/MRI-stabilization process may only apply
for those flow conditions in which there is no normal
flow on either one or both radial boundaries of the
system.  \par
On the other hand,
for a similar magnetized fluid configuration described by
periodic boundaries which therefore permits channel modes
to exist, modifications
of the shear appears to further destabilize the channel mode.
If one were to consider a flow configuration
(e.g. the local Keplerian profile
in a shearing box threaded by a constant background field) that
admits an unstable channel mode, then there seems to be no
modification of the shear with zero mean that would saturate the
growth of the channel mode \emph{since all such modifications
further enhances its destabilizing influence}.  The emergence of a pattern state
leading to saturation like the one envisioned by the hypothesis
would seem to be
out of the question under these circumstances.
Moreover this suggests that
the tendency for angular momentum transport reduction with P$_{{\rm m}}$,
as indicated by
 numerical experiments of the shearing box,
 is due to some other process or processes (which are not discussed here). \par
 These issues deserve further investigation as the conclusions reached
 here were done so utilizing very simple functional forms for the
 shear profile $Q$.  Namely, the analysis performed in the previous
 sections either assumed weak values of $Q$, in order to use
 singular perturbation theory techniques, or that $Q$ is described
 by delta-functions (see also below).  Use of these functions, which greatly facilitate analysis,
 have offered some insights into the nature of the responses of these
 modes.  Of course, further investigation would require considering more physically
 realizable profiles and to check the robustness of the results they yield
 against the ones offered by use of these more simplified forms.
 As such the results reached here can serve as guideposts for further
 enquiries along these lines.
 \par
In the following some reflections are presented
on other issues pertaining
to the implications of the calculations performed here.
 \subsection{On the use of delta-functions}
 A comparison between the responses of a single delta-function shear
 profile on an infinite domain (Section \ref{single_defect})
 and a top-hat profile for $Q$ (Section \ref{symmetric_shear_step}) show that
 the results of the two configurations are qualitatively similar. Although
 the symmetric step function allows for more normal modes to exist
 as its radial extent is increased - a feature which is missed
 if one uses the delta-function prescription - the qualitative
 flavor of the response is captured
 by its use nevertheless.
 \par
 Another deficiency found in using the defect prescription
 is that for the top-hat profile studied in
Section \ref{symmetric_shear_step}
there are both even and odd parity normal
 modes possible whereas only the even-parity mode is admitted
 in the defect model.
 The odd-parity mode
 \emph{has no counterpart} in the single defect model.  It was also shown that the odd-parity
 mode may exist if the radial domain of the symmetric shear step
 is large enough as (\ref{min_L_criterion_odd_parity_mode})
 indicates that the minimum size of the domain required
 for this mode to exist has the proportionality $L\sim1/\sqrt{Q_0}$.
 For small values of the defect amplitude the minimum size of the
 shear step required for this odd-parity mode to exist
 is so large that one does not expect that this should play
 any physically relevant role in the dynamics if one is interested
 in the response to narrow regions of altered shear.  It is with some
 confidence to suppose that
 it is reasonable to represent the dynamical response
 of slender regions of modified shear through the use of delta-function
 defects and this is the motivation and justification
 for its use in Sections 5.5 and 5.6.

\subsection{Channel modes}
In the small $Q$ theory of Section \ref{small_Q_theory},
it was shown that channel modes exist
for periodic boundary conditions and for conditions in which
the total pressure perturbations are zero at the two boundaries.
Conversely, if either one of the two boundaries force the
velocity perturbation to be zero there, then there
is no channel mode permitted as a normal mode solution.  In
light of the previous results it is important that
the boundary
conditions appropriate for disks be properly ascertained
in order to assess and better understand how activity in
them are driven.\par
Channel modes in the classical limit have no radial structure.
In Section \ref{ShearDefectPeriodicDomain} the development
of this mode was studied as a function of the
defect amplitude $Q_0$ and it is shown that though the
channel mode develops some amount of radial structure,
as evinced by the non-zero value of its wavenumber $\kappa_{_{00}}$,
its tendency to be unstable survives and is even enhanced.
In Regev \& Umurhan (2008) it was postulated that the channel mode
might disappear as a natural mode of the system if some amount
of radial symmetry breaking were introduced into the governing
equations of motion, for instance, in the form of a radially
varying shear profile.  However, the results of this section
show that the classical channel mode indeed persists in its
existence despite the introduction of a shear profile
which deviates from the background constant shear state with
zero average over some length scale.  These conclusions
are consistent with
similar results regarding the nature of the MRI in more global
contexts (Curry et al. 1994).
\subsection{On the existence and number of localized normal modes}
In Sections \ref{single_defect} and \ref{symmetric_shear_step}
the existence of localized modes was examined.  Localization
is understood in this case to correspond to modal disturbances
which show exponential decay as the radial coordinate $x\rightarrow \pm\infty$.
Thus, by construction, incoming/outgoing waves are excluded from
consideration. \footnote{Given the mathematical structure of the ode
describing axisymmetric disturbances, the radial eigenfunctions
that are solutions have either strictly oscillatory or
exponential radial profiles.  Thus the possibility of
decaying oscillations is ruled out.}  The existence of localized normal modes depends on
the relative sign of the shear defect/step profile $Q_0$:  HMI modes
exist if the shear in the defect/top-hat region is stronger than the background
($Q_0 > 0$) while HI modes exist if
the shear is relatively weakened ($Q_0 < 0$)
\footnote{The mode existence dependency upon the sign of $Q_0$ was checked for other shear
profiles (for instance, by replacing the top-hat with a truncated parabolic profile).  Aside
from differences in the magnitude of the temporal response and the multiplicity of modes allowed,
the aformentioned sign dependency still holds.}
-  summarized here as
\beqa
& & Q_0 >0 \longleftrightarrow {\rm hydromagnetic} \  {\rm inertial} \ {\rm disturbances}, \nonumber \\
& & Q_0 <0 \longleftrightarrow {\rm hydro-inertial} \ {\rm disturbances}. \nonumber
\eeqa
Given that the former mode type leads to the
MRI, it means that if the shear in the defect zone is weakened, then there are no
localized normal modes which can go unstable as only hydro-inertial modes are supported
(which happen to be stable except for very extreme circumstances).
Of course, as an initial value problem,
it does not mean that there
are no hydromagnetic inertial modes permitted but, instead, the mode probably
has some type of algebraic temporal growth/decay associated with it which
cannot be described by the usual normal mode approach.
This needs to be further examined by studying the associated initial-value problem.
Similar features are known to exist for vertically localized MRI modes
in disks (Liverts \& Mond, 2009) in which the response of
the initial value problem  can exhibit exponential growth with an
amplitude supporting some algebraic time-dependence.
Growth can initially be very small and it can take a very long
time for a given disturbance to reach the same amplitude as
expected for its counterpart normal mode.  A conjecture is
that the meaning
of the absence of normal modes in the problem considered here
may have similar attributes to the features associated with
the initial-value problem examined by
Liverts \& Mond (2009).
\par
Irrespective of this outcome it
should be remembered that the dichotomy observed here for
modes on an infinite radial domain
is a consequence of the imposition of exponential decay of the disturbances.
This distinction disappears once waves are allowed to enter and exit from
the ``infinite" boundaries.\par
  {Lastly it is also noteworthy that for the localized problems
considered here, when normal modes are permitted they usually
come as a finite set.  This is in contrast to other localized flow problems
considered elsewhere in the literature.  For example, in the problem of
compact rotating magnetized jets in an infinite medium considered by Bodo et al. (1989) a countably infinite number of normal-modes are predicted for that system.  This has
some origin in the uniformity of both the rotation and magnetic field profiles
of the jet considered in that study.  By contrast, there are examples in geophysical flows in which the number of normal mode disturbances are discrete and finite.  A recent example
can be found in the work of Griffiths (2008) where the stability of inertial waves
are studied in a stratified rotating flow with strong horizontal shear.  For strongly
peaked forms of the potential vorticity only a finite set of normal modes
are predicted for the system.  Such finiteness of the number of normal
modes permitted is not so unusual in systems in which background
states, like the shear or stratification, have strongly peaked functional forms like
they are in the cases studied in this work.}

\subsection{Stabilization in general and an interpretive tool}
The absolute minimum condition that must be
met for any normal mode of the system to be unstable
in a configuration with constant shear $q_0$ is
given by the Velikhov criterion (\ref{velikhov_condition}):
if the square of the Alfven frequency ($\omega_{az}^2$) is greater than
the shear by $2\Omega_0^2 q_0$, then none of the
HMI modes of the system
can be unstable.  For individual modes in the same
system with some amount
of radial structure the criterion is given by (\ref{criterion_for_instability_classic_mri}),
where $\beta_n^2$ is the square of the radial wavenumber of the disturbance.
As can be seen, for fixed values of $\omega_{az}^2$ there are two ways
in which stability may be promoted - either by weakening the shear or by
increasing the radial wavenumber.  In the problem where the shear
is constant, these quantities are set from the outset as parameters. \par
\begin{figure}
\begin{center}
\leavevmode \epsfysize=8.15cm
\epsfbox{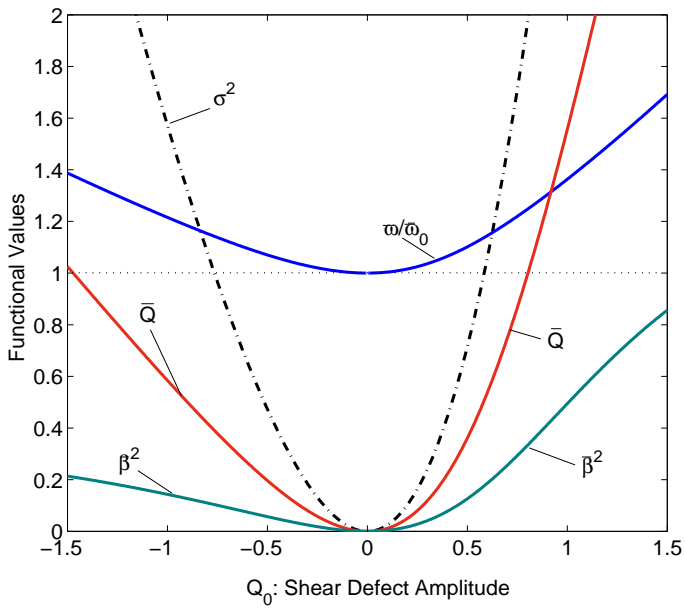}
\end{center}
\caption{The behavior of the unstable channel mode for the problem
with periodic boundary conditions examined in Section
\ref{ShearDefectPeriodicDomain}:
$kL = \pi$ and $\omega_{az}^2
= 2q_0$ with $q_0 = 3/2$.
Plotted is the temporal response, $\sigma^2$, together with
its corresponding effective wavenumber and shear $\bar \beta$ and $\bar q =
q_0 + \bar Q$ ($\bar Q$ plotted), and the stability measure for this
mode $\varpi/\varpi_0$.  All values of the shear defect $Q_0$ correspond
to increasing the effective shear to above the background value $\bar Q(Q_0) > 0$.
This increase in $\bar q$ always overpowers the stabilizing influence
of an increased $\bar\beta^2$.
Note, the values of $\sigma^2$
have been scaled by an arbitrary factor in order to facilitate clear comparison.}
\label{section6plot1}
\end{figure}

  {At the end of Section 4 there is a series of arguments
for arbitrary shear profiles which lead to the identity found in (\ref{genA}),
which is like a generalized eigenvalue condition. } This relationship
is the analog of the dispersion condition of the classical limit (uniform $q$)
contained in (\ref{dispersion_condition_uniform_q}).\footnote{The classical
limit is obtained when the replacements $\bar\beta \rightarrow \beta_n$
and $q\rightarrow q_0$ are made.}   { Care must be adopted before interpreting this
expression as an eigenvalue condition since $\sigma$ appears implicitly in
the expressions for $\bar q$ and $\bar\beta$.}  Nonetheless,  a parallel analysis
shows that stability is promoted if
\beq
\varpi \equiv \omega_{az}^2 - \frac{2\Omega_0^2 \bar q k^2}{\bar\beta^2 + k^2} > 0,
\label{interpretive_tool}
\eeq
where the effective wavenumber $\bar\beta$ and  shear $\bar q$,
given in (\ref{genB},\ref{genC}) are,
\[
\bar\beta^2 \equiv
 \frac{\int_{{\cal D}}|\partial_x\psi|^2 dx}
 {\int_{{\cal D}}|\psi|^2 dx}, \qquad
 \bar q \equiv q_0 + \frac{\int_{{\cal D}}Q|\psi|^2 dx}
 {\int_{{\cal D}}|\psi|^2 dx}.
\]
${\cal D}$ is the domain and $\psi$ is the radial eigenmode
of the disturbance in question.  The stability condition
for general $q$ may be rationalized in the same
way as the criterion is understood for the classical case
with the replacements, $\beta_n \rightarrow \bar\beta$
and $ q_0 \rightarrow \bar q$.    {With these aforementioned observations
and caveats in mind,
one may  use (\ref{interpretive_tool}) as an interpretive tool to aid in understanding
the reasons for stability/instability in a particular problem.}
Thus, stability is promoted
 if (i) for fixed effective wavenumber,  the effective shear
 is reduced or (ii) for fixed effective shear the effective
 wavenumber is increased, or in more general terms (iii) if
 the combination results in an overall reduction of the
 quotient
 \[
 \frac{\bar q}{\bar\beta^2 + k^2}.
 \]
 If one considers the results concerning the problem with
 periodic boundaries in Section \ref{ShearDefectPeriodicDomain},
 then the persistent instability
 expected for channel modes can be understood in terms
 of this criterion.  In the classical limit where the
 defect amplitude is zero then $\bar\beta$ for the channel
 mode is zero.  As the defect amplitude is raised $\bar\beta$
 increases suggesting that this effect has a stabilizing influence,
 e.g. see Figure \ref{section6plot1}.
 However,  as the defect amplitude is raised away from zero
 the effective shear increases no matter what the sign of $Q_0$
 and, to effect, this increase always sufficiently overpowers
 the increase in $\bar\beta^2$ so that the end result is further instability.
 In the same figure is plotted the quantity
 \beq
 \frac{\varpi}{\varpi_0} =
 \frac{\bar q}{q_0}\frac{\beta_n^2 + k^2}{\bar\beta^2 + k^2},
 \eeq
 which, for the generalized Velikhov criterion,
 is a measure of the ratio of the destabilizing
 term for the shear profile $q$ compared to its nominal value when the shear is only $q_0$:
 for ${\varpi}/{\varpi_0} > 1$ one expects enhanced destabilization
 against the classical mode
 while for ${\varpi}/{\varpi_0} < 1$ comes with enhanced stabilization.
 In Figure \ref{section6plot1} this quantity is plotted for one of the channel modes
 discussed in \ref{ShearDefectPeriodicDomain}.  Noting that
 in this case $\beta_n = 0$ (i.e. that its radial wavenumber is
 zero in the ideal case) the radial profile of the mode develops as
 the amplitude of the shear defect increases no matter what its sign.
 The stabilizing rise in $\bar\beta$ is accompanied by the destabilizing
 increase of the effective shear, that is to say $\bar Q > 0$,
 so that the quotient of their respective influences
 results in ${\varpi}/{\varpi_0} > 1$, indicating instability.
 \par
\begin{figure}
\begin{center}
\leavevmode \epsfysize=8.15cm
\epsfbox{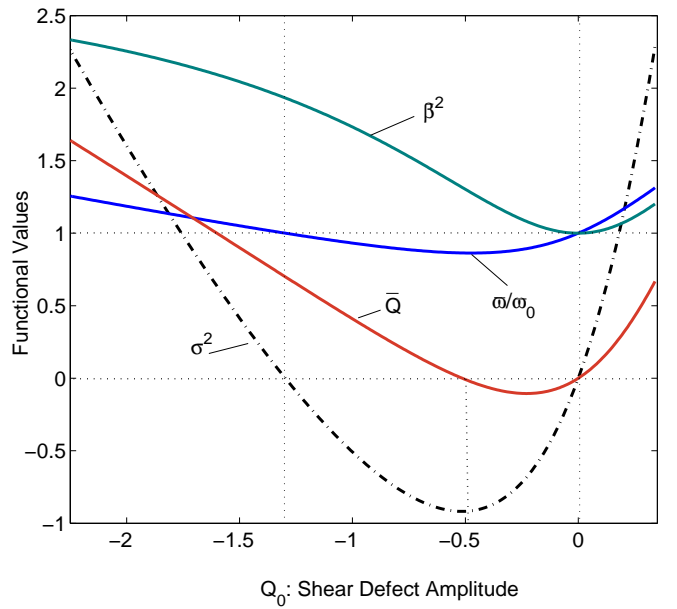}
\end{center}
\caption{The behavior of the least stable HMI mode for the problem
with channel walls examined in Section \ref{finite_domain_single_defect_channel_walls}:
$kL = \pi$ and $\omega_{az}^2
= 2q_0/(k^2 + \beta_1^2)$ with $\beta_1 = \pi/L$ and $q_0 = 3/2$.
Plotted is the temporal response, $\sigma^2$, together with
its corresponding effective wavenumber and shear $\bar \beta$ and $\bar q =
q_0 + \bar Q$ ($\bar Q$ plotted), and the stability measure for this
mode $\varpi/\varpi_0$.  Stability occurs for $Q_0 \in (-1.32, 0)$.  In this
window the effective shear is less than the background value, ($\bar q - q_0 =
\bar Q < 0$) for only
part of this range, i.e. for $Q_0 \in (-0.5,0)$.  Stabilization in the
range $Q_0 \in (-1.32,-0.5)$
is achieved by a sufficient
\emph{increase} of the effective wavenumber so that the
overall effect upon the stabilization parameter $\varpi/\varpi_0$
is for it to be less than one.  $\sigma^2$
has been similarly scaled as in the previous figure.}
\label{section6plot2}
\end{figure}

 Similarly plotted in Figure \ref{section6plot2} is  the influence
 $Q_0$ has upon $\bar\beta$,
 $\bar q$, and the stability measure $\varpi/\varpi_0$, for the
 single defect problem on a finite domain studied
 in Section \ref{finite_domain_single_defect_channel_walls}.
 The presence of walls filters out the channel mode and the
 resulting flow potentially can support shear profiles which can stabilize
 the least stable mode.  The Figure shows
 that within the range of (negative) values of the defect amplitude
 $Q_0$ for which the mode does not exponentially grow/decay
 the effective shear is less than the background shear state
 only for part of the stable range.  It means to say, then,
 in that range of parameter values in which the effective
 shear is greater than the background shear, yet there is still
 stability of the mode, stabilization is brought through
 an increased effective radial wavenumber which over-compensates the
 increased destabilization brought about by an increase
 in the effective shear.

\subsection{Final reflections and an implication}
The fundamental basis of this study is the assumption that in low magnetic Prandtl number
flows for model environments like that considered in this study, the only noticeable outcome
of the development of the MRI is to alter the basic shear profile.  All other quantities,
such as magnetic fields and the radial and vertical velocities, have saturated profiles
which scale as some positive power of P$_{\rm m}$ so that for P$_{\rm m} \ll 1$
their contributions become negligible.  The basis of this assumed trend derives from the
theoretical calculations of laboratory setups discussed in the Introduction.  It is therefore assumed here that
a similar trend \emph{may} appear for configurations in shearing boxes with periodic radial
boundary conditions as well.  The perspective taken here with regards to those configurations is that
if this is indeed the case, then one can (in principle) test the stability of a wide variety
of shear profiles which show deviations (constrained to have a zero average over some length scale)
 from the basic Keplerian profile - configurations which are akin to
 those predicted from the calculations
 of laboratory setups.    If a stability calculation shows that there exists a shear profile
 which is stable to these axisymmetric disturbances, then the shear-modification/mode-saturation
 hypothesis should become a serious candidate explanation.\par
 However,
 the calculations done in this study indicate that for periodic shearing box environments
there is always an instability of the basic channel mode
no matter what shear profile is assumed
(satisfying the aforementioned
constraints).   Unlike the situation
for laboratory setups, where saturation can be achieved by altering the basic shear profile,
saturation (if present) in shearing box environments
with periodic boundary conditions \emph{is likely not due to this mechanism}.\par
This further implies that the nonlinear response of systems supporting the
MRI stongly depends upon the boundary conditions employed which is
not so surprising as many physical systems in Nature exhibit this
sensitivity to their boundaries.
Thus it seems to this author that some care must be taken before
one equates the results of laboratory experiments to those physically
related/analogous systems for which
the experiments are meant to represent
(cf. Ji et al. 2006).  This cautionary
note is not to diminish the value of one over the other,
rather, it is intended to emphasize that there are many subtle and, at times, conflicting
features between such systems that must be clearly understood before any
firm conclusions are reached.
\section{Acknowledgements}
The author thanks Marek Abramowicz and the organizers of the Asymptotic Methods in Accretion Disk Theory
workshop at CAMK (May/June 2009) where the impetus for this work originated.  The author
also thanks James Cho, Lancelot Kao, Paola Rebusco, Edgar Knobloch and
Oded Regev for their generous support and fruitful conversations during
the course of this study.

\appendix
\section{Proof that $\sigma^2$ is never complex}\label{Integral_Argument}
(\ref{MHD_psi_eqn}) is rewritten into the form
\[
\partial_x^2 \psi -
k^2\left(1+\frac{\omega_0^2}{\sigma^2 + \omega_{{az}}^2}
- \frac{4\Omega_0^2 \omega_{_{az}}^2}{(\sigma^2 + \omega_{{az}}^2)^2}
-\frac{2\Omega_0^2 Q}{\sigma^2 + \omega_{{az}}^2}\right)\psi = 0,
\]
where $\omega_0^2 = 2\Omega_0^2(2-q_0)$.  The nature of $\sigma^2$
will be assessed by supposing that
\[
\sigma^2 +  \omega_{{az}}^2 = a + i b
\]
where $a,b \in$ Reals.  It will be shown below that $b$ must be zero.\par
The boundary conditions on $\psi$ are one of the following
possibilities: (i) $\psi$ goes to zero on the domain boundary
$\partial {\cal D}$ provided the latter is finite, (ii)  $\psi$ and
 its derivative go to zero on the domain boundary if the boundary tends to
 infinite distances, (iii) $\psi$ is periodic on a length scale $L$
in the domain ${\cal D}$.
Replacing $\sigma^2 +  \omega_{{az}}^2$ with $a + ib$ in the equation
for $\psi$, followed by multiplication by the complex conjugate of $\psi$,
(i.e. $\psi^*$), integrating the result and applying the boundary conditions gives
\beq
I_1 + i I_2 = 0, \label{integral_constraint}
\eeq
where
\beqa
& & I_1 =
-\int_{{\cal D}} |\partial_x \psi|^2 dx \nonumber \\
& &  -k^2\int_{{\cal D}}{
\left[1 + a\frac{\omega_0^2 - 2\Omega_0^2 Q}{a^2 + b^2}
-4\Omega^2 \omega_{{az}}^2\frac{a^2-b^2}{a^2 + b^2}
\right]|\psi|^2dx}
, \ \ \
\eeqa
and
\beqa
& & I_2 =
b k^2 \int_{{\cal D}}{
\left[\frac{\omega_0^2 - 2\Omega_0^2 Q}{a^2 + b^2}
- \frac{4\Omega_0^2 \omega_{{az}}^2 2a}{(a^2 + b^2)^2}\right]|\psi|^2 dx}.
\eeqa
It must be that both $I_1$ and $I_2$ are zero if (\ref{integral_constraint}) is to be satisfied.
Supposing that $b\neq 0$ then the relationship implied by $I_2=0$, is that
\[
\int_{{\cal D}}{
\left[\frac{\omega_0^2 - 2\Omega_0^2 Q}{a^2 + b^2}\right]|\psi|^2 dx}
=
\int_{{\cal D}}{
 \left[\frac{4\Omega_0^2 \omega_{{az}}^2 2a}{(a^2 + b^2)^2}\right]|\psi|^2 dx}.
\]
This result used in the expression for $I_1$ lead to
\beqa
& & I_1 =
-\int_{{\cal D}} |\partial_x \psi|^2 dx
 -k^2\left[1
+\frac{4\Omega^2 \omega_{{az}}^2}{a^2 + b^2}
\right]\int_{{\cal D}}{
|\psi|^2dx}.
\eeqa
However all the quantities in the above expression are always positive and, therefore,
$I_1$ can never be zero.  Consequently $b$ must be zero
which, in turn, means that $\sigma^2$ can only be real.\par
With $b=0$ and restoring the definition of $a$ the equation $I_1 = 0$ appears now
as
\beqa
& & 0=
-\int_{{\cal D}} |\partial_x \psi|^2 dx \nonumber \\
& &  -k^2\int_{{\cal D}}{
\left[1 + \frac{\omega_0^2 - 2\Omega_0^2 Q}{\sigma^2 +  \omega_{{az}}^2}
-\frac{4\Omega^2 \omega_{{az}}^2}{(\sigma^2 +  \omega_{{az}}^2)^2}
\right]|\psi|^2dx}.
\eeqa
Dividing the above expression by $\int_{{\cal D}} |\psi|^2 dx$
and utilizing the definition of $\kappa_0^2$ given in (\ref{kappa_0_def})
leads to the expressions appearing in (\ref{genA}-\ref{genC}).

\end{document}